\documentclass[letterpaper,journal]{IEEEtran}
\usepackage{amsmath,amsfonts}
\usepackage{algorithmic}
\usepackage{algorithm}
\usepackage{array}
\usepackage{amssymb}
\usepackage{balance}
\usepackage{makecell}
\usepackage[caption=false,font=normalsize,labelfont=sf,textfont=sf]{subfig}
\usepackage{textcomp}

\usepackage{url}
\usepackage{verbatim}
\usepackage{graphicx}
\usepackage{xcolor}
\usepackage{cite}
\usepackage{booktabs}
\usepackage{hyperref}
\usepackage{placeins}
\usepackage{subcaption}
\usepackage{float}
\usepackage{multirow}
\usepackage{siunitx}
\begin{document}

\title{MDIRNET: Multi-Degradation Image Restoration Network via Deep Unfolding}

\author{Talha Nadeem,
	Arslan Majal,
	Muhammad Tahir~\IEEEmembership{Senior Member,~IEEE},
	and Khurram Ali~\IEEEmembership{Senior Member,~IEEE}%
	\thanks{Talha Nadeem, Arslan Majal, and Muhammad Tahir are with the Department of Electrical Engineering, Lahore University of Management Sciences (LUMS), Lahore 54792, Pakistan (e-mail: talha.nadeem@lums.edu.pk; majal@wisc.edu; tahir@lums.edu.pk).}%
	\thanks{Khurram Ali is with the Department of Electrical and Computer Engineering, COMSATS University, Islamabad, Pakistan (e-mail: khurram.ali@cuilahore.edu.pk).}%
	\thanks{\textcopyright~2026 IEEE. Personal use of this material is permitted. Permission from IEEE must be obtained for all other uses, in any current or future media, including reprinting/republishing this material for advertising or promotional purposes, creating new collective works, for resale or redistribution to servers or lists, or reuse of any copyrighted component of this work in other works.}%
}

\maketitle

\begin{abstract}
	Real images often exhibit unknown and mixed degradations, making restoration substantially more challenging than single-task image restoration because multiple distortion types interact within the same observation. Consequently, existing methods often rely on prior knowledge of the degradation type or separate task-specific models, which may oversmooth fine structures or leave residual artifacts, motivating a compact model-driven alternative. We propose the Multi-Degradation Image Restoration Network (MDIRNET), a unified framework that combines a model-driven low-rank prior with end-to-end learning. Here, unified refers to joint training on three degradation types: noise, rain, and blur. A single MDIRNET model restores all three without requiring task-specific models, modules, or branches at inference. The low-rank prior exploits the redundancy and compact structure of natural image patches. To identify this underlying low-dimensional representation, we formalize restoration via Orthogonal Variational PCA (OVPCA) and translate its iterative inference into a deep unfolding network. To handle spatially non-uniform corruption and local content variability, we further introduce a learnable patch-partitioning strategy and a lightweight dynamic rank-allocation module that predicts the appropriate subspace dimension for each region. Spatially adaptive reconstruction refinement is performed using a supervised attention module. Extensive experiments on standard denoising, deblurring, and deraining benchmarks show that MDIRNET achieves competitive or superior performance over strong baselines across most metrics, while controlled mixed-degradation experiments demonstrate consistent performance across the evaluated synthetic degradation combinations. The code is available at \url{https://github.com/ScholarForge/mdirnet.git}.
	
\end{abstract}


\section{Introduction}\label{sec:intro}
Image restoration aims to recover a clean, high-quality image from a degraded version caused by factors such as noise, motion blur, or rain streaks. It is a long-standing challenge in computer vision with broad applications in autonomous driving, medical imaging, remote sensing, and more \cite{ding2021perceptual,wu2020medical,shen2022coupling}. In practice, images are often affected by multiple forms of degradation simultaneously, which can severely degrade the performance of downstream tasks that require clean visual inputs. Restoring images under mixed degradations is substantially more difficult than handling a single type of corruption, as the restoration model must account for different distortion characteristics in a single framework.\\
Most restoration methods target a single known degradation, making them impractical for unknown or compound distortions that may otherwise require multiple task-specific models. Recent multi-degradation methods~\cite{li2022all} partly address this issue but often rely on separate branches or known corruption types, limiting their flexibility.\\
Traditional image restoration relies on model-based priors, including PCA for low-rank denoising~\cite{greenacre2022principal}, total variation~\cite{wu2021adaptive}, and wavelet filtering~\cite{huang2022winnet}. Although interpretable, these methods often struggle with complex or unknown forms of degradation. Deep learning has substantially improved denoising~\cite{liu2020connecting,jiang2022deep}, deblurring~\cite{dong2021deep,li2023image}, and deraining~\cite{yang2020single,wang2020model}, but most models remain specialized to individual distortions and generalize poorly beyond their training domain. This motivates a unified restoration model that handles diverse degradations without prior knowledge of the corruption type.\\
We propose a unified multi-degradation restoration architecture based on Orthogonal Variational PCA (OVPCA)~\cite{vsmidl2007bayesian}, a probabilistic PCA framework with orthogonality constraints and variational inference. We unfold OVPCA into a learnable model-driven restoration network. While deep-unfolded OVPCA provides a model-driven low-rank backbone, it is insufficient on its own for multi-degradation restoration. Spatially varying degradations make fixed-patch processing suboptimal, varying local complexity makes a fixed-rank prior restrictive, and structured residual errors may not be fully captured by the low-rank prior. To address these limitations, MDIRNET incorporates a Patch Partitioning Module (PPM), a Dynamic Rank Allocation Module (DRAM), and a lightweight Supervised Attention Module (SAM). Together, these modules extend the model-driven low-rank backbone into a locally adaptive restoration framework, where patch allocation, the retained subspace dimension, and reconstruction refinement are jointly learned within the complete architecture. The main contributions of this work include:
\begin{itemize}
	\item We formulate a deep unfolding network that translates the iterative optimization of OVPCA into a sequence of learnable layers. The resulting Deep Unfolded (DU-OVPCA) core provides a compact model-driven low-rank backbone around which the adaptive components of MDIRNET are jointly designed.
	
	\item We integrate a learnable PPM module that adapts patch sampling to local image characteristics and a DRAM that predicts the retained rank of each patch group. Together, these modules enable content-adaptive patch processing and subspace allocation within the DU-OVPCA framework.
	
	\item  We incorporate a lightweight SAM module that compares the restored image's features with the original input to estimate regions where further refinement is beneficial. SAM produces an attention map that blends the restored and original inputs, thereby refining the final output by emphasizing regions that require additional correction.
	
	\item Extensive experiments show that MDIRNET achieves competitive or superior quality across denoising, deraining, and deblurring benchmarks while maintaining a compact model size of 0.68\,M parameters. Ablation studies quantify the contribution of each component.
\end{itemize}
The rest of the paper is organized as follows. Section 2 reviews related work on image restoration methods. Section 3 details our proposed MDIRNET framework, including the OVPCA-based formulation, deep unfolding architecture, and the relevant modules. Section 4 presents experimental results on individual and combined degradation scenarios, and Section 5 discusses ablation studies. Finally, Section 6 concludes the paper with insights and future directions.

\section{Related Work} Image restoration has been extensively studied for various types of degradation. Here, we review key developments in image denoising, deblurring, and deraining, followed by recent efforts toward unified restoration across multiple degradations.
\subsection{Image Denoising} 
Classical denoising exploits non-local self-similarity and sparse or low-rank priors. BM3D~\cite{cao2023improved} uses collaborative filtering of similar patches, while low-rank~\cite{dong2012nonlocal} and multi-scale models~\cite{su2020multi} regularize reconstruction. Deep learning further improves performance: DnCNN~\cite{murali2020image} learns residual noise, FFDNet~\cite{zhang2018ffdnet} uses a tunable noise-level map, and RIDNet~\cite{zhuo2019ridnet} employs attention for sharper restoration. Blind denoisers such as CBDNet~\cite{guo2019toward} jointly estimate and remove unknown noise, while self-supervised strategies~\cite{moran2020noisier2noise,krull2019noise2void} enable training without clean targets. Overall, deep methods significantly outperform classical ones, especially for complex or real-world non-Gaussian noise~\cite{liu2022deep}.

\subsection{Image Deblurring} Classical deblurring methods formulate blind deconvolution with explicit blur priors~\cite{levin2011understanding,levin2009understanding}, but struggle with spatially varying or large-kernel blurs. Non-blind techniques such as Wiener filtering~\cite{biswas2015deblurring} and Richardson--Lucy~\cite{yang2014novel} assume a known kernel, limiting practical applicability. Deep learning enables end-to-end mappings from blurred to sharp images: multi-scale CNNs~\cite{nah2017deep,tao2018scale} progressively restore detail at increasing resolutions, while GAN-based methods like DeblurGAN-v2~\cite{kupyn2019deblurgan} produce perceptually sharp outputs. Recent work exploits hierarchical patch processing~\cite{zhang2019deep}, multi-temporal recurrence~\cite{park2020multi}, and transformer architectures for modeling long-range dependencies. 

\subsection{Image Deraining} Early deraining methods model the rainy image as a superposition of a clean background and a rain layer, using sparsity priors~\cite{li2017single}, dictionary learning~\cite{xiao2018single}, or low-rank approximation~\cite{luo2020single} to separate the two. Non-local patch grouping~\cite{li2018non,kim2013single} provides moderate improvement but struggles with spatially varied heavy rain. Deep CNN-based approaches now dominate: RESCAN~\cite{li2018recurrent} removes rain progressively across recurrent stages, SPANet~\cite{li2020spatial} captures multi-scale streak information via spatial pyramid attention, and DRD-Net~\cite{deng2019drd} handles heavy rain without oversmoothing. Semi-supervised strategies such as UMRL~\cite{yasarla2019uncertainty} and SIRR~\cite{wei2019semi} further improve generalization to real rain conditions.

\subsection{Unified Image Restoration Frameworks} 
Recent all-in-one methods include AirNet~\cite{li2022all}, which learns degradation-invariant representations but may overlook localized degradation, and MPRNet~\cite{zamir2021multi}, which uses a multi-stage architecture with higher computational cost. Unified restoration remains challenging due to computational complexity, spatially varying degradations, and extensive mixed-degradation training. Recent methods address these issues through efficient architectures~\cite{cui2023irnext} and degradation-aware representations or conditioning~\cite{tang2025degradation,zhang2025perceive}. Motivated by these limitations, MDIRNET combines OVPCA-based unfolding with adaptive patch partitioning, dynamic rank allocation, and supervised attention for unified multi-degradation restoration.\\
Image restoration is directly relevant to instrumentation and measurement, as image quality affects measurement fidelity in sensor-based pipelines. In autonomous driving, rain and motion blur can degrade object detection and distance estimation. In medical imaging, acquisition noise in MRI and CT can affect diagnostic measurements, while atmospheric blur and sensor noise in remote sensing can impair land-cover classification and target detection. Prior studies have addressed image denoising and enhancement in such measurement contexts~\cite{zhang2023considering,yang2009multifocus}. MDIRNET targets these scenarios with a unified model that handles multiple simultaneous degradations without prior knowledge of their types or severities, thereby improving the reliability of downstream measurements.\\
A key motivation for adopting low-rank modeling in this multi-degradation context is the well-established observation that clean natural image patches $\mathbf{x} \in \mathbb{R}^p$ lie near a low-dimensional manifold $\mathcal{M}_r \subset \mathbb{R}^p$ with $\text{rank}(\mathbf{M}) = r \ll p$ \cite{elad2006image,mairal2009online}. Any degradation type $k$ distorts this geometry: the observed patch matrix satisfies $\mathbf{D} = \mathbf{X} + \mathbf{E}_k$. While $\mathbf{E}_k$ can inflate the effective rank, the singular-value perturbation bound (Observation 1) motivates low-rank regularization under bounded perturbations, providing an inductive bias rather than a guarantee. The OVPCA framework, therefore, provides a degradation-agnostic inductive bias: it recovers the clean low-dimensional subspace and discards out-of-subspace energy through a variational shrinkage step, without requiring any prior knowledge of the specific corruption type. For convolutional blur, the low-rank prior serves as a regularizer rather than a complete physical model, while PPM and SAM compensate for information beyond rank-$r$ projection. This makes low-rank modeling suitable for unified multi-degradation restoration.

\section{Proposed Method}
\subsection{Preliminaries: OVPCA and DU-OVPCA} 
\begin{figure}[h]
	\centering
	\includegraphics[width=0.99\linewidth]{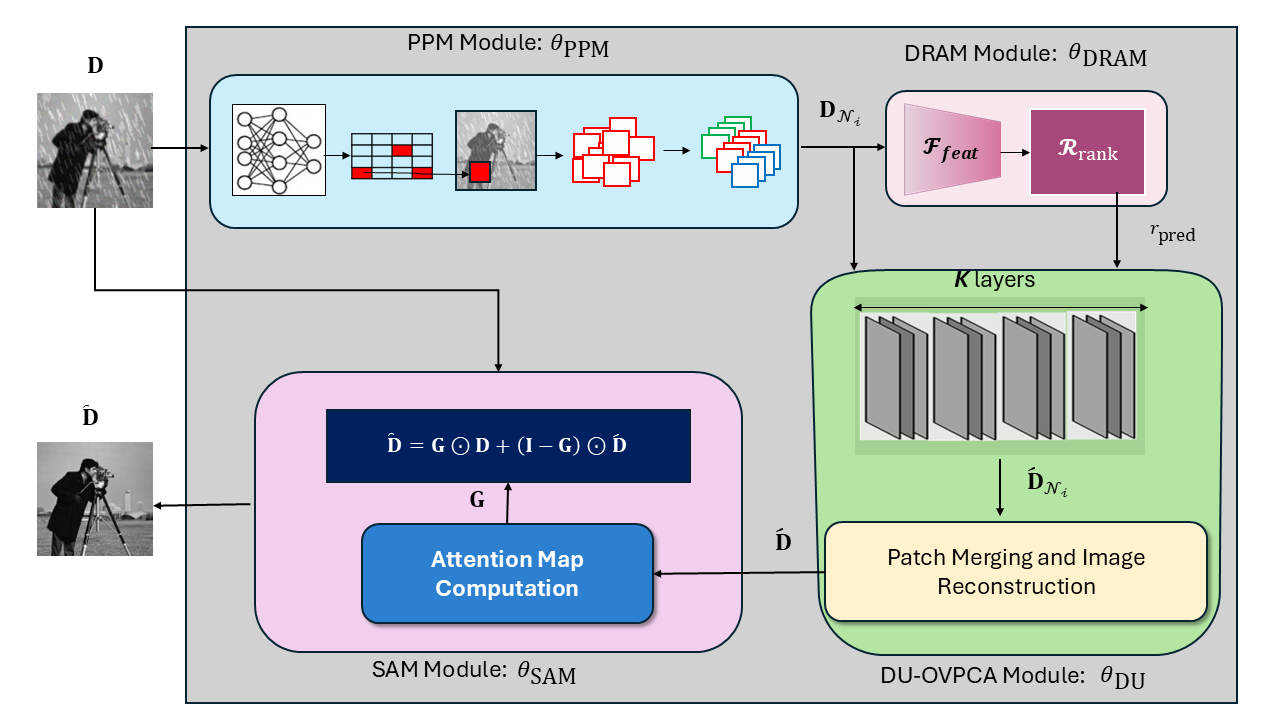}
	\caption{Proposed MDIRNET framework with the DU-OVPCA core and PPM, DRAM, and SAM modules.
	}
	\label{fig:main}
\end{figure}

A central modeling assumption of our framework, grounded in the manifold-geometry argument of Section \eqref{sec:intro}, is that clean image patches lie near a low-rank subspace. At the same time, any corruption type distorts this geometry. This is a patch-level statistical prior, not a universal physical law, and is consistent with prior works on dictionary learning and sparse representation \cite{elad2006image,mairal2009online}. Traditional PCA seeks a low-dimensional linear subspace that captures the data. For an image, we can vectorize it and consider it as lying in an $r$-dimensional subspace spanned by principal components. Let $\mathbf{D}\in\mathbb{R}^{p\times n}$ denote the degraded patch matrix. Classical PCA approximates $\mathbf{D} \approx \mathbf{A}\mathbf{\Sigma} \mathbf{X}^{T} + \mathbf{E}$, where $\mathbf{A} \in \mathbb{R}^{p \times r}$, $\mathbf{X} \in \mathbb{R}^{n \times r}$ with orthonormality constraints $\mathbf{A}^{T}\mathbf{A} = \mathbf{I}_r$, $\mathbf{X}^{T}\mathbf{X} = \mathbf{I}_r$, and $\mathbf{\Sigma} = \operatorname{diag}(\sigma_1, \ldots, \sigma_r)$ contains the associated singular values. The factorization is unique up to column-sign ambiguity. However, PCA's rigid linear subspace assumption is inadequate for complex degradations.\\
OVPCA~\cite{vsmidl2007bayesian} extends PCA into a probabilistic framework, treating the clean image as a latent variable:
\begin{equation}
	\mathbf{D} = \mathbf{A} \mathbf{L} \mathbf{X}^T + \mathbf{E}
\end{equation}
where $\mathbf{A}$, $\mathbf{X}$ are orthonormal and $\mathbf{L}\in\mathbb{R}^{r\times r}$ is diagonal. OVPCA places priors on $\mathbf{A}$, $\mathbf{X}$, and noise precision $\omega$ for variational inference. Assuming a Gaussian likelihood for the data: $p(\mathbf{D}|\mathbf{A},\mathbf{L},\mathbf{X},\omega) = \mathcal{N}(\mathbf{A}\mathbf{L}\mathbf{X}^T,\ \omega^{-1} I)$, and placing priors $p(\mathbf{A}) = \mathcal{U}(\mathcal{S}_{p,r})$, $p(\mathbf{X}) = \mathcal{U}(\mathcal{S}_{n,r})$ (uniform over the Stiefel manifold, enforcing orthogonality), and $p(\omega) \propto \omega^{-1}$ (Jeffreys' non-informative prior), the posterior is $p(\mathbf{A},\mathbf{L},\mathbf{X},\omega|\mathbf{D},r) \propto p(\mathbf{D}|\mathbf{A},\mathbf{L},\mathbf{X},\omega) \cdot p(\mathbf{A}) \cdot p(\mathbf{X}) \cdot p(\omega)$. The MAP objective is:
	\begin{equation}
		(\hat{\mathbf{A}},\hat{\mathbf{L}},\hat{\mathbf{X}},\hat{\omega}) = \arg \max_{\mathbf{A},\mathbf{L},\mathbf{X},\omega} p(\mathbf{A},\mathbf{L},\mathbf{X},\omega|\mathbf{D},r).
	\end{equation}
Since the exact posterior is intractable due to integration over the Stiefel manifold, Variational Bayes (VB) approximates it using the factored distribution $q(\boldsymbol{\theta}) = q(\mathbf{A})q(\mathbf{X})q(\mathbf{L})q(\omega)$ by minimising the KL divergence $\text{KL}(q(\boldsymbol{\theta}) \| p(\boldsymbol{\theta}|\mathbf{D}))$, equivalently maximising the Evidence Lower Bound (ELBO). The MAP objective defines \emph{what} is optimized, whereas VB defines \emph{how}; the iterative updates below are coordinate-ascent steps that monotonically increase the ELBO \cite{blei2017variational}.\\
The result is an iterative algorithm that alternately updates $\mathbf{A}, \mathbf{L}, \mathbf{X}$ and $\omega$, to maximize the posterior. Assuming rank-$r$ Singular Value Decomposition (SVD) of data matrix as $\mathbf{D} \xrightarrow{\text{SVD}} \{ \mathbf{U}_r, \mathbf{V}_r, \mathbf{l}_r \}, \quad \hat{\omega}^{(0)} = \frac{pn}{\| \mathbf{l}_D - \mathbf{l}_r \|_2^2}$, the iterative updates of OVPCA are given as  \cite{vsmidl2007bayesian}
\begin{equation*}
	\begin{cases} 
		\mathbf{k}_{A,X}^{(t)} &= \mathcal{G} \left( \hat{\omega}^{(t-1)}, \mathbf{l}_r, \mathbf{k}_{X,A}^{(t-1)}, \hat{\mathbf{l}}^{(t-1)} \right) \\ 
		\mathbf{m}_{\mathbf{l}}^{(t-1)} &=\mathbf{k}_X^{(t-1)} \odot \mathbf{l}_r \odot \mathbf{k}_A^{(t-1)}\\
		\hat{\mathbf{l}}^{(t)} &= \mathbf{m}_{\mathbf{l}}^{(t-1)} + s^{(t-1)} \Psi(\mathbf{m}_{\mathbf{l}}^{(t-1)}, s^{(t-1)}) \\ s^{(t)} &= \left( \hat{\omega}^{(t-1)} \right)^{-1/2}, \quad 
		\hat{\omega}^{(t)} = \frac{pn}{\mathcal{R}(\mathbf{k}_A, \mathbf{k}_X, \hat{\mathbf{l}})}\\
		\hat{\mathbf{D}} &= (\mathbf{U}_r \mathbf{K}_A) \cdot \text{diag}(\hat{\mathbf{l}}) \cdot (\mathbf{V}_r \mathbf{K}_X)^T
	\end{cases}
\end{equation*}
The rank-$r$ truncated SVD is $\mathbf{D}_r = \mathbf{U}_r \Sigma_r \mathbf{V}_r^T$, 
where $\mathbf{U}_r, \mathbf{V}_r$ have orthonormal columns and $\Sigma_r = \operatorname{diag}(\sigma_1, \ldots, \sigma_r)$.
Concretely, $\mathcal{G}$ computes element-wise scaling corrections $k_{A, i}^{(t)} = \bigl(1 + \hat{\omega}^{(t-1)} (l_{r, i} \cdot k_{X, i}^{(t-1)} \cdot \hat{l}_i^{(t-1)})^2\bigr)^{-1/2}$, the closed-form VB update for the expected basis-column norms~\cite{vsmidl2007bayesian}; in the unfolded network its inputs are outputs of previous learnable layers. The shrinkage operator $\Psi(m_l, s) = \max(|m_l|-s,0)\cdot\operatorname{sign}(m_l) - m_l$ soft-thresholds singular values with a learned threshold $s=(\hat{\omega})^{-1/2}$, suppressing noise components while preserving signal. The scaled basis $\hat{\mathbf{A}}=\mathbf{U}_r\operatorname{diag}(\mathbf{k}_A)$ is non-orthonormal in general: $\hat{\mathbf{A}}^T\hat{\mathbf{A}} = \operatorname{diag}(\mathbf{k}_A)^2 \neq I$ unless $|k_i|=1$ for all $i$. After unfolding, the ELBO is no longer optimized; instead, we optimize $L_{rec}$. Its structure nevertheless provides an inductive bias that encourages the learnable updates to follow coordinate-ascent VB, where $\mathbf{l}_D$ is the vector of singular values of $\mathbf{D}$ and $\mathcal{R}$ computes the residual energy.\\
\textbf{Observation 1} (Low-Rank Prior Motivation).
	\textit{Let $\mathbf{X}\in\mathbb{R}^{p\times n}$ have rank $r$ and
		$\mathbf{D}=\mathbf{X}+\mathbf{E}$, where $\|\mathbf{E}\|_2=\epsilon$.
		By the singular-value perturbation bound,
		\begin{equation}
			|\sigma_i(\mathbf{D})-\sigma_i(\mathbf{X})|
			\leq \epsilon,
		\end{equation}
		and since $\sigma_i(\mathbf{X})=0$ for $i>r$,
		\begin{equation}
			\sigma_r(\mathbf{D})\geq\sigma_r(\mathbf{X})-\epsilon,
			\qquad
			\sigma_i(\mathbf{D})\leq\epsilon,\quad i>r.
		\end{equation}
Thus, when $\epsilon<\sigma_r(\mathbf{X})$, the $r$th singular value remains nonzero, while those beyond rank $r$ are bounded by $\epsilon$, motivating low-rank regularization under bounded perturbations.} This observation does not imply rank inflation or guarantee clean-image recovery; outside the bounded-perturbation regime, the low-rank prior is an empirical inductive bias. Unlike robust PCA~\cite{bouwmans2018applications}, which typically assumes sparse corruption, or sparse coding~\cite{elad2006image}, which requires an explicit dictionary, OVPCA provides MDIRNET's model-driven low-rank backbone, whose effectiveness is validated empirically.\\
Each DU-OVPCA layer costs $O(Gm^2CTr)$. At $256{\times}256$ with $m{=}8$, $T{=}8$, $r{=}32$, $G{=}512$: ${\sim}0.04$~GFLOPs/layer, ${\sim}0.24$~GFLOPs for $K{=}6$ --- a fraction of the total 14.0~GFLOPs.\\
\begin{figure}[h]
    \centering
    \includegraphics[width=0.9\linewidth]{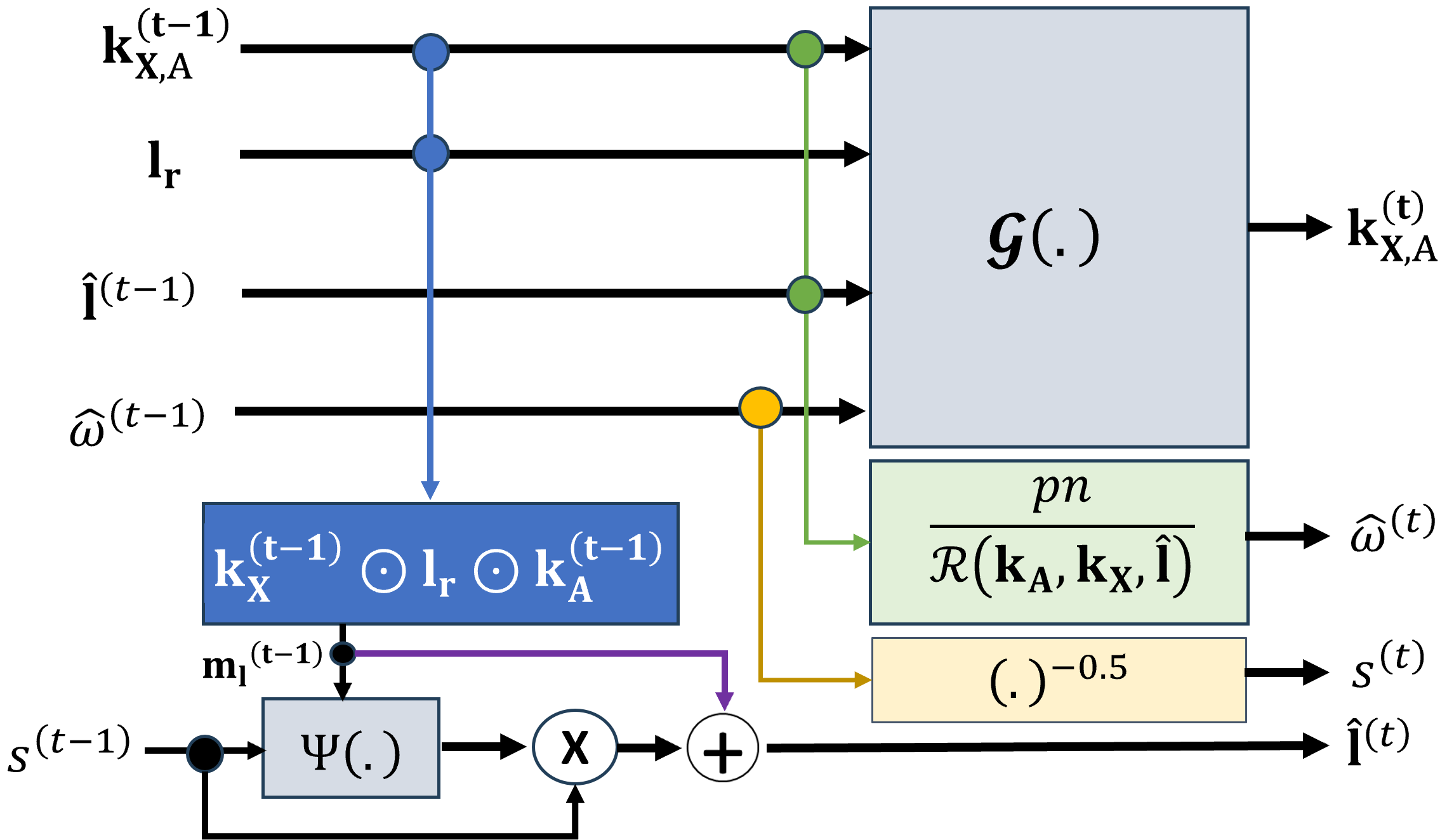}
    \caption{Single layer of DU-OVPCA architecture}
    \label{fig:enter-label}
\end{figure}
The iterative OVPCA solution can be unfolded into a deep network by mapping each iteration to a network layer and training all layers jointly end-to-end. The resulting network, termed Deep Unfolding OVPCA (DU-OVPCA), refines the clean image representation over $K$ layers corresponding to $K$ OVPCA iterations, as illustrated in Fig.~\ref{fig:enter-label}. This architecture incorporates restoration priors while enabling discriminative parameter learning and preserving the structured inductive bias of the variational updates. The VB coordinate-ascent updates of OVPCA satisfy the ELBO monotonicity property: each iteration non-decreasingly increases $\mathcal{L}(q) = \mathbb{E}_q[\log p(\mathbf{D},\boldsymbol{\theta})] - \mathbb{E}_q[\log q(\boldsymbol{\theta})]$, which is bounded above by $\log p(\mathbf{D}|r)$, guaranteeing convergence to a local stationary point \cite{blei2017variational}. Empirically, our layer-depth ablation shows that performance saturates at $K=6$ unfolding layers, with negligible additional improvement at $K=8$. Although this theoretical guarantee does not directly extend to the end-to-end trained unfolded network, the original ELBO monotonicity provides the architectural motivation for the unfolded update sequence.
The architecture of MDIRNET is shown in Fig.~\ref{fig:main}. Below, we detail each module.

\subsubsection{Patch Partitioning Module (PPM)} Real-world degradations are often spatially non-uniform, making uniform restoration suboptimal. We address this using a learnable PPM module that partitions the image into content-aware patches instead of a fixed grid, redistributing sampling locations according to learned spatial and image-content characteristics.\\
PPM is implemented as a small CNN that takes the degraded image $\mathbf{D}\in\mathbb{R}^{H\times W \times C}$ (height $H$, width $W$, $C$  channels) and outputs a flow field $\mathbf{\Delta}\in\mathbb{R}^{H\times W \times 2}$. This flow field can be interpreted as offsets to a regular grid of patch sampling coordinates. Formally, let $\mathbf{G}_{\text{base}}(x,y)$ denote the normalized coordinates of pixel $(x,y)$ in the image (ranging from $-1$ to $1$ across the width and height). The PPM predicts a dense offset $\mathbf{\Delta}(x,y) = ( \Delta_x, \Delta_y )$ at each location. We then define a deformed sampling grid:

\begin{equation}
	\mathbf{G}_{\text{deformed}}(x,y) = \mathbf{G}_{\text{base}}(x,y) + \mathbf{\Delta}(x,y)
\end{equation}
This grid slightly shifts the patch sampling positions. Using a standard differentiable sampling mechanism, we extract patches ${\mathcal{P}_i}$ from $\mathbf{D}$ centered at these shifted coordinates. In essence, the network learns to shift patch centers toward spatial regions that are emphasized by the restoration objective. When learned offsets from neighboring patch centers converge toward the same spatial region, i.e., when $\boldsymbol{\Delta}(x_1,y_1)$ and $\boldsymbol{\Delta}(x_2,y_2)$ point toward the same target, the effective sampling density increases at that location. This convergent-offset behavior emerges from end-to-end training rather than being guaranteed by the architecture.\\
In our implementation, PPM uses overlapping patches of size $m\times m$ ($m=8$), following common practice in low-rank patch-based restoration to balance structural consistency and computational efficiency. Adaptive multi-scale patch partitioning is left for future work. The PPM parameters $\mathbf{\theta}_{\text{PPM}}$ are trained end-to-end, meaning the network learns sampling patterns from the restoration training data. This contrasts with conventional patch-based methods, which use either a fixed grid or random sampling. By learning the sampling offsets, MDIRNET adaptively redistributes patch centers and can increase local sampling density in regions emphasized by the learned restoration strategy. This provides spatially adaptive patch processing in regions emphasized by the learned restoration objective.\\
After patch extraction, we also implement a patch grouping strategy. The set of patches ${\mathcal{P}_i}$ is partitioned into groups ${\mathcal{N}_i}$ based on spatial proximity. In our current design, for each reference patch $\mathcal{P}_i$ (centered at some image location), we collect its $k-1$ nearest neighboring patches in the spatial domain to form group $\mathcal{N}_i$. Grouping patches by spatial proximity efficiently approximates content-aware grouping, as PPM shifts neighboring patch centers toward spatially related regions emphasized by the learned sampling strategy. Compared with explicit non-local matching, this avoids quadratic complexity while enabling DU-OVPCA to exploit local structural redundancy for collaborative low-rank restoration.

\subsubsection{Dynamic Rank Allocation Module (DRAM)} A fixed rank $r$ is suboptimal: smooth regions need low rank while textured regions need high rank. DRAM is a lightweight meta-network (a convolutional feature extractor $\mathcal{F}_{\text{feat}}$ and an FC regression head $\mathcal{R}_{\text{rank}}$) that predicts the adaptive rank per patch group.  For each patch group
$\mathcal{N}_i=\{\mathbf{P}_1,\mathbf{P}_2,\dots,\mathbf{P}_T\}$,
where $T$ is the number of patches in the group, we first compute the
mean patch
\begin{equation}
	\bar{\mathbf{P}}_i=\frac{1}{T}\sum_{j=1}^{T}\mathbf{P}_j,
\end{equation}
which summarizes the overall complexity of the patch group while
maintaining negligible computational overhead. This mean patch
$\bar{\mathbf{P}}_i$ is then fed into the feature extractor
$\mathcal{F}_{\text{feat}}$, a lightweight CNN that produces a feature vector $\mathbf{v}_i$ characterizing the texture complexity of the patch group. The predicted rank is learned solely from the restoration objective, without explicit rank or computational-cost penalties. Thus, DRAM serves as a content-adaptive rank allocation mechanism rather than an explicit efficiency optimization. Mean pooling is adopted because it provides the best accuracy--efficiency trade-off among the evaluated aggregation strategies, as demonstrated by the ablation study in Section~V (Table~\ref{tab:ablation-quantitative}).
Intuitively, if the patches contain high-frequency patterns or strong edges, the feature will reflect that, whereas a flat patch group will produce a low-complexity feature. The feature vector $\mathbf{v}_i$ then goes into the rank prediction head $\mathcal{R}_{\text{rank}}$, which produces a scalar between 0 and 1. Let $\hat{r}_i \in [0,1]$ be this output.  We then scale and quantize $\hat{r}_i$ to an actual rank value as
\begin{equation} \label{rank_value}	
	r_{\text{pred}} = r_{\min} + 2 \cdot \operatorname{round}\!\left( \hat{r_{i}}\cdot \frac{r_{\max}-r_{\min}}{2}  \right)
\end{equation}
Here, $r_{\min}$ and $r_{\max}$ denote the minimum and maximum allowed ranks. Multiplication by 2 and rounding ensure an even integer rank. DRAM predicts an integer rank $r_{\text{pred}}$ for each patch group, which the DU-OVPCA module uses to limit the number of principal components during restoration. $\theta_{\text{DRAM}}$ denotes the module parameters.
Since $\operatorname{round}(\cdot)$ has zero gradient almost everywhere, end-to-end training of DRAM requires a Straight-Through Estimator (STE) \cite{bengio2013estimating}. During the forward pass, $r_{\text{pred}}$ is computed via rounding as in Eq.~\ref{rank_value}. During the backward pass, gradients are propagated as if rounding were an identity function: $\partial\mathcal{L}/\partial\hat{r}_i \approx \partial\mathcal{L}/\partial r_{\text{pred}}$. In PyTorch, this is implemented as \texttt{r\_pred~=~r\_cont~+~(r\_disc~-~r\_cont).detach()}, where \texttt{r\_cont} is the pre-rounding continuous value and \texttt{r\_disc} its rounded counterpart. This enables fully end-to-end training of DRAM jointly with the rest of MDIRNET.\\
During training, DRAM learns content-adaptive ranks through the restoration objective, assigning lower ranks to smooth regions and higher ranks to structurally complex regions. Since no explicit rank, FLOP, or latency penalty is used, DRAM does not directly optimize computational efficiency.

\subsubsection{Low-Rank Restoration via DU-OVPCA}

Once patches are extracted and ranks are determined for each patch group, the DU-OVPCA network performs core restoration as described earlier. For each patch group $\mathcal{N}_i$, we form a data matrix
$\mathbf{D}_{\mathcal{N}_i}\in\mathbb{R}^{(m^2C)\times T}$ by flattening each patch (of size $m\times m\times C$) into a $(m^2C)$-dimensional vector (concatenating color channels if $C>1$), and stacking these $T$ vectors as columns. We then seek a decomposition $ \mathbf{D}_{\mathcal{N}_i}\approx \mathbf{A}_i \mathbf{L}_i \mathbf{X}_i^{T} + \mathbf{E}_i$
where $\mathbf{A}_i\in\mathbb{R}^{(m^2C)\times r_i}$, $\mathbf{X}_i\in\mathbb{R}^{T\times r_i}$, and $\mathbf{L}_i$ is $r_i\times r_i$ diagonal with $r_i = r_{\text{pred}}$ given by DRAM for this group. The DU-OVPCA network runs for $K$ unfolding layers to estimate $\mathbf{A}_i, \mathbf{L}_i, \mathbf{X}_i$. At the end of this, we obtain a reconstructed clean patch group $ \widehat{\mathbf{D}}_{\mathcal{N}_i} = \mathbf{A}_i \mathbf{L}_i \mathbf{X}_i^{T}$ which consists of restored versions of each patch in the group. Due to the low-rank constraint, the network suppresses noise and outlier features that cannot be represented in the rank-$r_i$ subspace. Sharing the basis $\mathbf{A}_i$ across grouped patches further enforces consistency, enabling non-local restoration by sharing information among similar patches. We denote all parameters of the deep unfolding network by $\theta_{\text{DU}}$.\\
After processing all patch groups, the restored patches ${\widehat{\mathbf{P}}_j}$ are merged into the final image $\widehat{\mathbf{D}}$ by averaging their overlapping values. Specifically, each pixel $(x,y)$ is averaged over all patches covering that location:
\begin{equation}
	\widehat{\mathbf{D}}(x,y)=
	\frac{\sum_{j}\mathbb{I}\big[(x,y)\in \widehat{\mathbf{P}}_{\,j}\big]\cdot \widehat{\mathbf{P}}_{\,j}(x,y)}
	{\sum_{j}\mathbb{I}\big[(x,y)\in \widehat{\mathbf{P}}_{\,j}\big]}
\end{equation}
where $\mathbb{I}[\cdot]$ is an indicator function. This yields the initial restored image $\widehat{\mathbf{D}}$ after the DU-OVPCA restoration stage. While $\widehat{\mathbf{D}}$ removes most corruption, residual artifacts may persist in highly textured regions where the low-rank prior can oversmooth. This motivates the refinement module below.

\subsubsection{Supervised Attention Module (SAM)}
SAM refines the DU-OVPCA output $\widehat{\mathbf{D}}$ by providing spatially adaptive refinement through the combination of the restored output and information from the original observation $\mathbf{D}$. It extracts features from both images via parallel convolutions:
\begin{equation}
\mathbf{F}_{\text{deg}} = \operatorname{Conv}_1(\mathbf{D}), \quad
\mathbf{F}_{\text{res}} = \operatorname{Conv}_2(\widehat{\mathbf{D}})
\end{equation}
and produces a pixel-wise gate map $\mathbf{G}(x,y)=\sigma\!\left(\operatorname{Conv}_3\!\left([\mathbf{F}_{\text{deg}},\mathbf{F}_{\text{res}}]\right)\right) \in [0,1]$. The refined output is the convex combination:
\begin{equation}
\tilde{\mathbf{D}}(x,y) = \mathbf{G}(x,y) \cdot \mathbf{D}(x,y) + \bigl(1 - \mathbf{G}(x,y)\bigr) \cdot \hat{\mathbf{D}}(x,y)
\end{equation}
where $G\rightarrow0$ gives greater weight to the restored output $\hat{\mathbf{D}}$, while $G\rightarrow1$ gives greater weight to the original degraded observation $\mathbf{D}$.
SAM adaptively combines restored and observed information through spatial gating. Since observations may contain degradation artifacts, the gate does not guarantee protection against local artifact reintroduction; its effectiveness is evaluated empirically. We use $\theta_{\text{SAM}}$ to indicate all parameters of the SAM module.
This SAM refinement is trained in a supervised manner using clean ground-truth images, enabling it to learn attention maps that minimize the final reconstruction error. SAM adaptively refines DU-OVPCA residuals by combining the original observation and restored representation to produce the final image $\tilde{D}$.

\subsection{Loss Function and Training} The entire MDIRNET framework is optimized end-to-end, with all parameters $\mathbf{\theta}_{\text{MDIRNET}} = \left\{\theta_{\text{PPM}}; \theta_{\text{DRAM}}; \theta_{\text{DU}}; \theta_{\text{SAM}}\right\}$ optimized jointly. We employ a hybrid $L_1$+$L_2$ reconstruction loss, where $L_1$ preserves edges and $L_2$ stabilises training:
\begin{equation}
\mathcal{L}_{\text{rec}}\left(\mathbf{\theta}_{\text{MDIRNET}}\right)
=
\left\lVert \tilde{\mathbf{D}} - \mathbf{D}_{\text{clean}} \right\rVert_1
+
\alpha \left\lVert \tilde{\mathbf{D}} - \mathbf{D}_{\text{clean}} \right\rVert_2^2
\end{equation}
We set $\alpha$ so that the $L_2$ term remains small relative to the $L_1$ term, reducing oversmoothing. We further introduce a patch-consistency regularizer to promote consistent restoration within each spatial group. Specifically, we penalize the intra-group variance of restored patch features to encourage consistency across overlapping patches:
\begin{equation}
\mathcal{L}_{\text{unc}}\left(\mathbf{\theta}_{\text{MDIRNET}}\right)
=
\mathbb{E}_i\Big[\log \left(1 + \,\operatorname{Var}\big(\{\widehat{\mathbf{P}}_{\,j}:\mathbf{P}_j\in\mathcal{N}_i\}\big)\,\right)\Big]
\end{equation}
i.e., the average variance of restored patch pixels within groups. A smooth transform such as $\log$ can reduce the effect of outliers. The regularizer controls intra-group variation, with $\lambda_{\mathrm{unc}}$ limiting excessive suppression of legitimate differences. Our total training loss is a weighted sum:
\begin{equation}
\mathcal{L}_{\text{total}}\left(\mathbf{\theta}_{\text{MDIRNET}}\right)
=
\lambda_{\text{rec}}\mathcal{L}_{\text{rec}}\left(\mathbf{\theta}_{\text{MDIRNET}}\right)
+
\lambda_{\text{unc}}\mathcal{L}_{\text{unc}}\left(\mathbf{\theta}_{\text{MDIRNET}}\right)
\end{equation}
with $\lambda_{\text{rec}}$ and $\lambda_{\text{unc}}$ balancing the terms. 

\section{Experiments and Results} We evaluate on standard benchmarks: BSD68~\cite{martin2001database} (68 images, Gaussian noise $\sigma{=}25,50$), BSD400~\cite{martin2001database} (400 images), WED~\cite{ma2016waterloo} (4744 images), DIV2K~\cite{Agustsson_2017_CVPR_Workshops} (1000 images), GoPro~\cite{nah2017deep} (blurred/sharp pairs), and Rain100L~\cite{yang2019joint} (200 training + 100 testing rainy/clean pairs).\\
We report PSNR (dB), SSIM, and average inference time (s). Baselines span classical methods (CBM3D~\cite{dabov2007color}), CNN-based approaches (DnCNN~\cite{zhang2017beyond}, FFDNet~\cite{zhang2018ffdnet}, BRDNet~\cite{tian2020image}), and recent unified restoration frameworks (AirNet~\cite{li2022all}, Restormer~\cite{zamir2022restormer}, DA-RCOT~\cite{tang2025degradation}, Perceive-IR~\cite{zhang2025perceive}, MPRNet~\cite{zamir2021multi}, DeAirRTUR~\cite{wu2025debiased}, AWRaCLe~\cite{rajagopalan2025awracle}). Task-specific baselines are listed in the respective tables.

\subsection{Implementation Details}
The proposed framework was implemented in PyTorch~2.0.1 on an NVIDIA RTX~3090 GPU (24~GB VRAM) with CUDA~11.8. For optimization, we used the AdamW optimizer with $\beta_1{=}0.9$, $\beta_2{=}0.999$, weight decay $10^{-4}$, and an initial learning rate of $10^{-4}$ decayed via cosine annealing to $10^{-7}$. The training batch size was set to 4, with input patches of size $64{\times}64$ randomly cropped during training. Data augmentation, including random rotation and horizontal flipping, was applied. The network was trained for 100 epochs with model weights saved every 10 epochs. Key hyperparameters are: $K{=}6$ unfolding layers, $r_{\max}{=}32$,
$r_{\min}{=}4$, loss weights $\lambda_{\text{rec}}{=}1.0$,
$\lambda_{\text{unc}}{=}0.1$ (selected via ablation,
Table~\ref{tab:lambda_unc}), and $\alpha{=}0.5$. For our implementations, inference time is the median of 100 forward passes after 10 warm-ups (batch size 1). Literature results follow their published protocols, while reproduced/retrained methods follow Table~\ref{tab:baseline_protocol}. MDIRNET and the blind baselines use only the degraded image at inference, whereas DA-RCOT additionally receives degradation-type information and is therefore reported as a degradation-aware reference rather than a directly equivalent comparison.\\
The PPM module uses 3 convolutional layers with 96 channels each. The DRAM feature extractor has a single convolutional layer with 32 channels followed by two fully connected layers ($32\rightarrow32\rightarrow1$). The SAM module uses 64-channel convolutions. These channel widths keep the model below 1~M parameters: the DU-OVPCA core performs restoration in singular-value space, while the auxiliary modules act as lightweight adapters. Table~\ref{tab:module_breakdown} provides the parameter breakdown.
\begin{table}[t]
	\centering
	\caption{Per-module parameter, FLOP, and runtime breakdown at $256\times256$ input.}
	\label{tab:module_breakdown}
	\scriptsize
	\setlength{\tabcolsep}{1.5pt}
	
	\resizebox{\columnwidth}{!}{%
		\begin{tabular}{|l|p{3.2cm}|c|c|c|c|c|}
			\hline
			\textbf{Module} & \textbf{Architecture} & \textbf{Params} & \textbf{GFLOPs} & \textbf{\% GFLOPs}
			 & \textbf{Runtime (ms)} & \textbf{\% Runtime} \\
			\hline
			
			PPM &
			3 Conv: $3\!\to\!96\!\to\!96\!\to\!2$ ($3\times3$)
			& 0.60M & 11.8 & 84.3\% & 92.0 & 47.2\% \\
			\hline
			
			DU-OVPCA &
			6 DU-OVPCA layers ($r\le32$)
			& 0.02M & 0.24 & 1.7\% & 73.0 & 37.4\% \\
			\hline
			
			DRAM &
			Conv $3\!\to\!32$ + FC $32\!\to\!32\!\to\!1$
			& 0.01M & 0.01 & 0.1\% & 6.0 & 3.1\% \\
			\hline
			
			SAM &
			3 Conv: $3\!\to\!64$, $3\!\to\!64$, $128\!\to\!1$ ($3\times3$)
			& 0.05M & 1.95 & 13.9\% & 24.0 & 12.3\% \\
			\hline
			
			\multicolumn{2}{|l|}{\textbf{Total}} &
			\textbf{0.68M} &
			\textbf{14.0} &
			\textbf{100\%} &
			\textbf{195.0} &
			\textbf{100\%} \\
			\hline
			
		\end{tabular}%
	}
\end{table}
While PPM accounts for 84.3\% of GFLOPs, its highly parallelized convolutions contribute only 47.2\% of wall-clock runtime. In contrast, DU-OVPCA accounts for just 1.7\% of GFLOPs but 37.4\% of runtime due to the sequential nature of SVD operations, highlighting that GFLOPs alone do not reliably predict runtime (Table~\ref{tab:module_breakdown}). For the backbone comparison in Table~\ref{tab:backbone}, all methods were trained and evaluated under identical datasets, optimizer, learning-rate schedule, batch size, unfolding stages, and evaluation protocol to ensure a fair comparison.\\
Table~\ref{tab:baseline_protocol} summarizes baseline provenance, including training settings, retraining status, result sources, and inference inputs. Unless marked as retrained, baseline results are from the cited sources and follow their reported protocols. Table~\ref{tab:backbone} evaluates all variants under a unified protocol.
\begin{table*}[t]
	\centering
	\caption{Training/inference protocols of major baselines. 'O' and 'R' denote original and retrained results, respectively.
	}
	\label{tab:baseline_protocol}
	\setlength{\tabcolsep}{2pt}
	\renewcommand{\arraystretch}{0.72}
	\resizebox{\textwidth}{!}{
		\begin{tabular}{|l|p{7.2cm}|c|p{3.4cm}|}
			\hline
			\textbf{Baseline} &
			\textbf{Training data / setting} &
			\textbf{Src.} &
			\textbf{Inference input} \\
			\hline
			
			CBM3D, DnCNN, IRCNN, FFDNet &
			Task-specific Gaussian denoising &
			O~\cite{dabov2007color,zhang2017beyond,zhang2017learning,zhang2018ffdnet} &
			Degraded image; noise level when required \\
			\hline
			
			BRDNet &
			BSD400/WED-based denoising &
			O~\cite{tian2020image} &
			Degraded image + noise level \\
			\hline
			
			AirNet &
			BSD400/WED/Urban100 (denoising); Train100L/Rain100L (deraining); RESIDE-OTS (dehazing)
			 &
			O~\cite{li2022all} &
			Degraded image \\
			\hline
			
			UMRL, SIRR &
			Task-specific semi-supervised deraining &
			O~\cite{yasarla2019uncertainty,wei2019semi} &
			Degraded image \\
			\hline
			
			MSPFN &
			13,712 clean/rain image pairs &
			O~\cite{jiang2020multi} &
			Degraded image \\
			\hline
			
			LPNet &
			1,800 heavy-rain + 200 light-rain images; 3M generated patches &
			R~\cite{fu2019lightweight} &
			Degraded image \\
			\hline
			
			MFFDNet &
			Four synthetic + two real-world deraining datasets &
			O~\cite{tao2024mffdnet} &
			Degraded image \\
			\hline
			
			MPRNet &
			Task-specific denoising, deraining, and deblurring &
			O~\cite{zamir2021multi} &
			Degraded image \\
			\hline
			
			Restormer &
			Rain13K (deraining); GoPro (motion deblurring) &
			O~\cite{zamir2022restormer} &
			Degraded image \\
			\hline
			
			Perceive-IR &
			Noise+Haze+Rain; extended to Blur+Low-light
			 &
			O~\cite{zhang2025perceive} &
			Degraded image \\
			\hline
			
			DA-RCOT &
			All-in-one 3D/5D degradation settings &
			O~\cite{tang2025degradation} &
			Degraded image + cues
			 \\
			\hline
			
			DeAirRTUR &
			AirNet + Task Uncertainty Regularization (TUR) &
			R~\cite{wu2025debiased} &
			Degraded image \\
			\hline
			
			AWRaCLe &
			All-weather restoration with contextual information &
			O~\cite{rajagopalan2025awracle} &
			Degraded image + context \\
			\hline
			
			DBGAN, MTRNN, DMPHN &
			GoPro-based motion deblurring &
			O~\cite{zhang2020deblurring,park2020multi,zhang2019deep} &
			Degraded image \\
			\hline
			
			MWFormer &
			All-weather/multi-weather restoration &
			O~\cite{zhu2024mwformer} &
			Degraded image \\
			\hline
			
			\textbf{MDIRNET} &
			\textbf{BSD400, WED, Rain100L, GoPro + unified settings
			} &
			\textbf{Ours} &
			\textbf{Degraded image only} \\
			\hline
			
		\end{tabular}
	}
\end{table*}

\subsection{Comparison for Single Degradation Scenarios}
Table~\ref{table1} reports denoising results on BSD68, Urban100, and DIV2K ($\sigma{=}25,50$). MDIRNET achieves an average PSNR of 31.82~dB / 0.898 SSIM, comparable to Perceive-IR (31.70~dB) while requiring shorter inference time (0.19~s vs.\ 0.23--0.33~s).
 These results show that MDIRNET provides competitive denoising accuracy with lower measured inference time under the evaluated settings. DA-RCOT is included as a degradation-aware reference and therefore operates under a different inference-information setting from MDIRNET; its numerical results should be interpreted accordingly.
\begin{table*}[h!]
	\centering
	\caption{Quantitative results of image denoising on BSD68, Urban100, DIV2K, and SIDD~\cite{abdelhamed2018high} datasets in terms of PSNR (dB), SSIM, average inference time (s), and average PSNR/SSIM.}
	\label{table1}
	\setlength{\tabcolsep}{3pt}
	\resizebox{\textwidth}{!}{
		\begin{tabular}{|l|c|c|c|c|c|c|c|c|c|c|c|}
			\hline
			\multirow{2}{*}{\textbf{Method}} 
			& \multicolumn{2}{c|}{\textbf{BSD68}} 
			& \multicolumn{2}{c|}{\textbf{Urban100}} 
			& \multicolumn{2}{c|}{\textbf{DIV2K}} 
			& \textbf{SIDD}
			& \multicolumn{3}{c|}{\textbf{Inference Time (s)}} 
			& \multirow{2}{*}{\textbf{Avg. PSNR/SSIM}} \\ \cline{2-11}
			& $\sigma=25$ & $\sigma=50$ 
			& $\sigma=25$ & $\sigma=50$ 
			& $\sigma=25$ & $\sigma=50$ 
			& Val
			& BSD68 & Urban100 & DIV2K & \\ \hline
			
			CBM3D \cite{dabov2007color}
			& 30.71/0.868 & 27.36/0.763 & 31.91/0.896 & 27.93/0.840 & 30.55/0.872 & 28.04/0.827 
			& 25.65/0.685
			& 0.31 & 1.20 & 9.00 & 28.88/0.822 \\ \hline
			
			DnCNN \cite{zhang2017beyond}
			& 31.23/0.880 & 27.92/0.788 & 30.81/0.893 & 27.59/0.833 & 31.00/0.888 & 28.10/0.833 
			& 36.98/0.920
			& 0.20 & 0.55 & 2.30 & 30.52/0.862 \\ \hline
			
			IRCNN \cite{zhang2017learning}
			& 31.18/0.880 & 27.88/0.789 & 31.26/0.899 & 27.70/0.840 & 30.90/0.883 & 27.90/0.821 
			& 36.40/0.912
			& 0.30 & 0.879 & 3.10 & 30.46/0.861 \\ \hline
			
			FFDNet \cite{zhang2018ffdnet}
			& 31.21/0.879 & 27.96/0.789 & 31.40/0.903 & 28.05/0.848 & 31.10/0.889 & 28.00/0.825 
			& 36.81/0.918
			& 0.28 & 0.70 & 2.80 & 30.65/0.864 \\ \hline
			
			BRDNet \cite{tian2020image}
			& 31.43/0.885 & 28.16/0.794 & 31.90/0.912 & 28.56/0.858 & 31.25/0.901 & 28.20/0.846 
			& 37.42/0.929
			& 0.22 & 0.65 & 2.60 & 30.99/0.875 \\ \hline
			
			AirNet \cite{li2022all}
			& 31.48/0.893 & 28.23/0.806 & 32.10/0.924 & 28.88/\textbf{0.870} & 31.45/0.908 & 28.40/0.852 
			& 38.84/0.951
			& 0.33 & 0.90 & 3.50 & 31.34/0.886 \\ \hline
			
			Restormer \cite{zamir2022restormer}
			& 31.78/0.909 & 28.59/0.822 & 33.02/0.927 & 28.33/0.851 & 31.55/0.912 & 28.43/0.853 
			& \textbf{40.02}/\textbf{0.960}
			& 0.33 & 1.05 & 4.10 & 31.67/0.891 \\ \hline
			
			DA-RCOT \cite{tang2025degradation}
			& 30.12/0.864 & 27.16/0.779 & 32.22/0.921 & 28.22/0.854 & 31.63/0.915 & 28.51/0.855 
			& 38.61/0.948
			& 0.23 & 0.68 & 2.70 & 30.92/0.877 \\ \hline
			
			Perceive-IR \cite{zhang2025perceive}
			& 31.74/0.908 & 28.53/0.821 & 32.55/\textbf{0.926} & 29.42/0.864 & \textbf{31.71}/0.922 & 28.55/0.857 
			& 39.41/0.956
			& 0.26 & 0.78 & 3.00 & 31.70/0.893 \\ \hline
			
			\textbf{MDIRNET} 
			& \textbf{31.84}/\textbf{0.920} & \textbf{28.64}/\textbf{0.826} 
			& \textbf{32.61}/\textbf{0.926} & \textbf{29.47}/0.867 
			& 31.69/\textbf{0.923} & \textbf{28.57}/\textbf{0.864} 
			& 39.91/\textbf{0.960}
			& \textbf{0.19} & \textbf{0.55} & \textbf{2.20} 
			& \textbf{31.82}/\textbf{0.898} \\ \hline
			
	\end{tabular}}
\end{table*}
On the Rain100L dataset (Table~\ref{table2}), earlier deraining baselines such as UMRL, SIRR, and MSPFN achieve only 27–32 dB, whereas recent methods like AirNet and MFFDNet achieve 34.90–35.61 dB, while DeAirRTUR reaches 38.04 dB. The DeAirRTUR result uses AirNet retrained with Task Uncertainty Regularization (TUR), rather than the original AirNet configuration. DeAirRTUR reports 38.04/0.983 on Rain100L, compared with 36.53/0.974 for MDIRNET, using a substantially smaller model and requiring less inference time (0.23~s vs.\ 0.33--0.44~s). This demonstrates competitive accuracy in heavy rain with substantially lower computational cost. The 1.63 dB difference reflects the empirical performance gap under their respective evaluation protocols; the specific sources of this difference are not isolated.
\begin{table*}[h!]
	\centering
	\caption{Results of image deraining on the Rain100L dataset in terms of PSNR (dB), SSIM, and inference time (s).}
	\label{table2}
	\setlength{\tabcolsep}{4pt}
	\begin{tabular}{|l|c|c|c|c|c|c|c|c|}
		\hline
		\textbf{Method} & \textbf{UMRL} \cite{yasarla2019uncertainty} & \textbf{SIRR} \cite{wei2019semi} & \textbf{MSPFN} \cite{jiang2020multi} & \textbf{LPNet} \cite{fu2019lightweight} & \textbf{AirNet} \cite{li2022all} & \textbf{MFFDNet} \cite{tao2024mffdnet} & \textbf{DeAirRTUR} \cite{wu2025debiased} & \textbf{MDIRNET} \\ \hline
		
		\textbf{PSNR/SSIM} 
		& 27.79/0.892 
		& 32.39/0.917 
		& 32.37/0.916 
		& 33.50/0.923 
		& 34.90/0.966 
		& 35.61/0.941 
		& \textbf{38.04}/\textbf{0.983} 
		& 36.53/0.974 \\ \hline
		
		\textbf{Inference Time (s)} 
		& 0.24 & 0.44 & 0.46 & 0.33 & 0.34 & 0.33 & 0.34 & \textbf{0.23} \\ \hline
		
	\end{tabular}
\end{table*}
Image deblurring experiments on the GoPro dataset (Table~\ref{table3}) further highlight the advantages of the proposed method. Earlier baselines such as DBGAN, MTRNN, and DMPHN do not exceed 31.2 dB PSNR, whereas more recent designs such as MPRNet, DA-RCOT, and MWFormer achieve 31.2-32.68 dB PSNR and SSIM values approaching 0.959. MDIRNET achieves 32.68~dB / 0.959 SSIM, matching MPRNet while requiring shorter inference time (0.23~s vs.\ 0.24~s). These results confirm MDIRNET's ability to restore sharp structures and textures in blurred images with accuracy and efficiency.\\
\begin{table*}[t]
	\centering
	\caption{Performance comparison on image denoising, deraining, and deblurring. Entries report PSNR (dB)/SSIM. Params: learnable parameters (M); GFLOPs: $256{\times}256$ input; Time: inference time (s). \textbf{Bold} indicates the best result.}
	\label{table4}
	\setlength{\tabcolsep}{3pt}
	\resizebox{\textwidth}{!}{
		\begin{tabular}{|c|l|c|c|c|c|c|c|c|c|c|c|}
			\hline
			\textbf{Type} & \textbf{Method}
			& \textbf{Params (M)} & \textbf{GFLOPs}
			& \textbf{BSD68 ($\sigma$=25)} & \textbf{BSD68 ($\sigma$=50)}
			& \textbf{Rain100L} & \textbf{GoPro}
			& \multicolumn{3}{c|}{\textbf{Inference Time (s)}}
			& \textbf{Avg. PSNR/SSIM} \\ \cline{9-11}
			& & & & & & & & BSD68 & Rain100L & GoPro & \\ \hline
			
			\multirow{5}{*}{\textbf{One-By-One}}
			& BRDNet \cite{tian2020image} & 1.12 & 9.50 & 31.43/0.885 & 28.16/0.794 & 32.15/0.915 & 31.10/0.942 & 0.22 & 0.28 & 0.32 & 30.71/0.884 \\ \cline{2-12}
			& LPNet \cite{fu2019lightweight} & 0.57 & 4.80 & 29.27/0.841 & 25.70/0.781 & 30.51/0.891 & 31.15/0.945 & 0.24 & 0.34 & 0.29 & 29.16/0.865 \\ \cline{2-12}
			& MPRNet \cite{zamir2021multi} & 15.74 & 588.0 & 31.43/0.910 & 28.10/0.801 & 34.72/0.968 & 32.66/\textbf{0.959} & 0.20 & 0.25 & 0.24 & 31.73/0.910 \\ \cline{2-12}
			& AirNet \cite{li2022all} & 8.93 & 75.6 & 31.48/0.893 & 28.23/0.806 & 34.90/0.966 & 31.20/0.940 & 0.25 & 0.34 & 0.32 & 31.45/0.901 \\ \cline{2-12}
			& \textbf{MDIRNET} & 0.68 & 14.0 & \textbf{31.84}/\textbf{0.920} & \textbf{28.64}/\textbf{0.826} & \textbf{36.53}/\textbf{0.974} & \textbf{32.68}/\textbf{0.959} & \textbf{0.19} & \textbf{0.23} & \textbf{0.23} & \textbf{32.42}/\textbf{0.920} \\
			\hline
			
			\multirow{8}{*}{\textbf{All-In-One}}
			& BRDNet \cite{tian2020image} & 1.12 & 9.50 & 29.45/0.889 & 27.00/0.812 & 31.48/0.902 & 28.60/0.864 & 0.23 & 0.30 & 0.29 & 29.13/0.867 \\ \cline{2-12}
			& LPNet \cite{fu2019lightweight} & 0.57 & 4.80 & 28.30/0.840 & 26.20/0.774 & 29.30/0.880 & 27.90/0.851 & 0.24 & 0.34 & 0.33 & 27.93/0.836 \\ \cline{2-12}
			& MPRNet \cite{zamir2021multi} & 15.74 & 588.0 & 30.89/0.901 & 27.90/0.835 & 34.20/0.933 & 30.60/0.939 & 0.20 & 0.26 & 0.24 & 30.90/0.902 \\ \cline{2-12}
			& AirNet \cite{li2022all} & 8.93 & 75.6 & 31.20/0.902 & 28.00/0.835 & 34.63/0.951 & 30.70/0.942 & 0.25 & 0.34 & 0.32 & 31.13/0.908 \\ \cline{2-12}
			& DA-RCOT \cite{tang2025degradation} & 18.50 & 87.4 & 31.47/0.909 & 28.21/0.842 & \textbf{38.36}/\textbf{0.983} & 30.33/0.890 & 0.23 & 0.28 & 0.27 & \textbf{32.09}/0.906 \\ \cline{2-12}
			& Perceive-IR \cite{zhang2025perceive} & 22.40 & 108.2 & 31.44/0.908 & 28.19/0.836 & 37.25/0.977 & 29.46/0.886 & 0.26 & 0.32 & 0.31 & 31.59/0.902 \\ \cline{2-12}
			& AWRaCLe \cite{rajagopalan2025awracle} & 14.80 & 68.5 & 31.33/0.906 & 28.13/0.834 & 35.42/0.961 & 30.15/0.882 & 0.28 & 0.34 & 0.32 & 31.26/0.896 \\ \cline{2-12}
			& \textbf{MDIRNET} & 0.68 & 14.0 & \textbf{31.56}/\textbf{0.913} & \textbf{28.33}/\textbf{0.845} & 35.21/0.958 & \textbf{31.34}/0.914 & \textbf{0.19} & \textbf{0.23} & \textbf{0.23} & 31.61/\textbf{0.908} \\
			\hline
			
	\end{tabular}}
\end{table*}
\begin{table*}[h!]
	\centering
\caption{Quantitative deblurring results on GoPro in terms of PSNR (dB), SSIM, and inference time (s).}
	\label{table3}
	\setlength{\tabcolsep}{4pt}
	\begin{tabular}{|l|c|c|c|c|c|c|c|}
		\hline
		\textbf{Method} & \textbf{DBGAN} \cite{zhang2020deblurring}  & \textbf{MTRNN} \cite{park2020multi} & \textbf{DMPHN} \cite{zhang2019deep}  & \textbf{MPRNet} \cite{zamir2021multi}  & \textbf{DA-RCOT}  \cite{tang2025degradation}   & \textbf{MWFormer} \cite{zhu2024mwformer}  & \textbf{MDIRNET} \\ \hline
		
		\textbf{PSNR/SSIM} 
		& 31.10 / 0.942 
		& 31.15 / 0.945 
		& 31.20 / 0.940 
		& 32.66 / \textbf{0.959} 
		& 31.41 / 0.936 
		& 31.92 / 0.951 
		& \textbf{32.68} / \textbf{0.959} \\ \hline
		
		\textbf{Inf. Time (s)} 
		& 0.32 & 0.29 & 0.28 & 0.24 & 0.27 & 0.31 & \textbf{0.23} \\ \hline
		
	\end{tabular}
\end{table*}
These results demonstrate that MDIRNET achieves competitive or superior performance across three representative restoration tasks while maintaining competitive inference time and a favorable efficiency--accuracy trade-off.

\subsection{Evaluation on Real-World Sensor Noise (SIDD)}
To further validate MDIRNET on real-world degradation, we evaluate on the SIDD benchmark~\cite{abdelhamed2018high}, which contains real smartphone sensor noise. As reported in Table~\ref{table1}, MDIRNET achieves 39.91~dB / 0.960 SSIM, demonstrating that the low-rank prior effectively handles real sensor noise as well as synthetic Gaussian noise. Figure~\ref{fig:single_noise} (Row~4) shows qualitative results. MDIRNET effectively suppresses real sensor noise while preserving color fidelity and fine structural detail, confirming that the low-rank prior generalizes beyond synthetic noise settings. After 25-epoch fine-tuning on DIV2K (Q10--Q40), MDIRNET achieves 29.51/0.816 and 31.72/0.883 on Classic5, and 29.35/0.812 and 31.66/0.881 on LIVE1 at Q10 and Q20, respectively. With only 0.68M parameters, it remains competitive with AirNet (8.93M).
\subsection{Quantitative Comparison for Multi-Degradation Scenarios}
\begin{figure}[t]
	\centering
	\includegraphics[width=\linewidth]{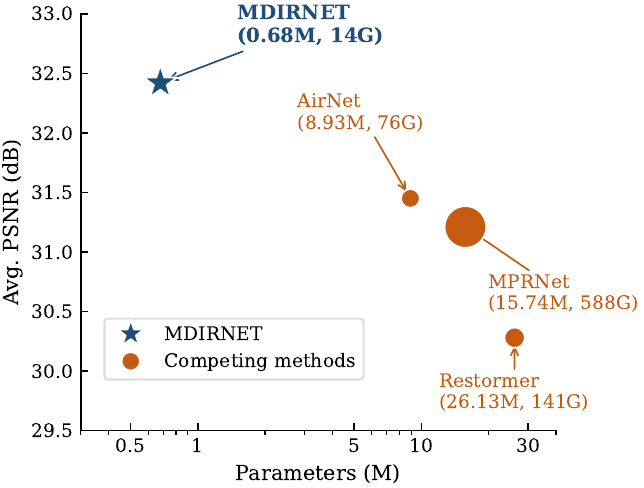}
	\caption{Parameter efficiency: Avg.\ PSNR vs.\ model size. Bubble area denotes GFLOPs.}
	\label{fig:pareto}
\end{figure}

\begin{figure*}[t]
	\centering
	\includegraphics[width=\textwidth]{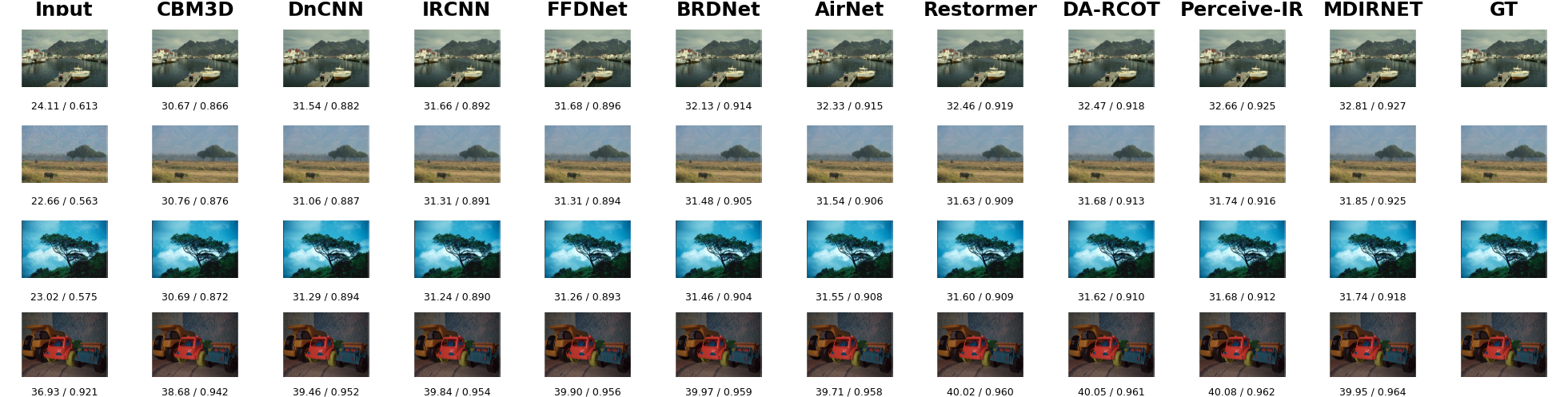}
	\caption{Qualitative comparison on noise degradation. Rows 1--3: synthetic Gaussian noise on BSD68. Row 4: real sensor noise from SIDD (\cite{abdelhamed2018high}). MDIRNET demonstrates superior denoising across both synthetic and real noise conditions.}
	\label{fig:single_noise}
\end{figure*}

\begin{figure*}[t]
	    \centering
	    \includegraphics[width=\textwidth]{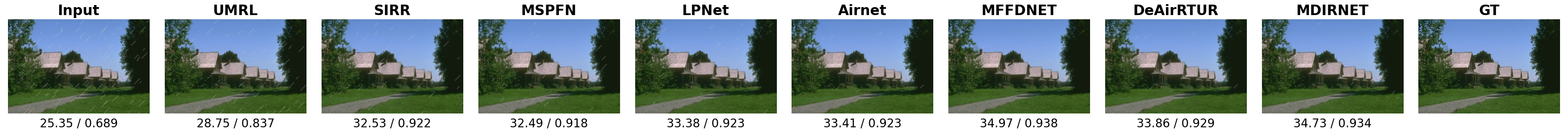}
	    \caption{Qualitative comparison on single rain degradation. MDIRNET effectively suppresses rain streaks while preserving background details and sharpness.}
	    \label{fig:single_rain}
	\end{figure*}

\begin{figure*}[t]
	    \centering
	    \includegraphics[width=\textwidth]{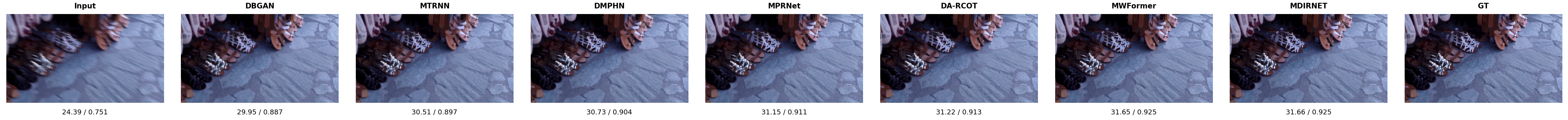}
	    \caption{Qualitative comparison on single-image deblurring. MDIRNET restores sharper edges and better structural consistency.}
	    \label{fig:single_blur}
	\end{figure*}

To evaluate MDIRNET under multi-degradation conditions, we evaluate images with simultaneous degradations, such as noise+blur and rain+noise, reflecting realistic mixed-degradation conditions. Mixed degradations are applied in a fixed order: blur$\rightarrow$noise for noise+blur, rain$\rightarrow$noise for noise+rain, and blur$\rightarrow$rain (50\% opacity) for rain+blur. The same order is used for training and evaluation. We compare MDIRNET with existing all-in-one restoration methods, including BRDNet and LPNet, as well as more recent multi-task networks. Results are reported in Table~\ref{tab:ablation} and Table~\ref{table4}.
\begin{table}[t]
	\centering
	\caption{Ablation study on combinations of degradations for \textbf{MDIRNET}. 
		``\checkmark'' denotes the presence of a degradation. Each entry is PSNR/SSIM. Best results are shown in boldface.}
	\label{tab:ablation}
	\resizebox{\linewidth}{!}{%
		\begin{tabular}{|c|c|c|c|c|c|}
			\hline
			\multicolumn{3}{|c|}{\textbf{Degradation}} &
			\textbf{Denoise (BSD68)} &
			\textbf{Derain (Rain100L)} &
			\textbf{Deblur (GoPro)} \\ \hline
			
			Noise & Rain & Blur &
			$\sigma=25$ / $\sigma=50$ &
			PSNR/SSIM &
			PSNR/SSIM \\ \hline
			
			\checkmark & -- & -- 
			& \textbf{31.84}/\textbf{0.920} \; / \; \textbf{28.64}/\textbf{0.826}
			& -- 
			& -- \\ \hline
			
			-- & \checkmark & -- 
			& -- 
			& \textbf{36.53}/\textbf{0.974}
			& -- \\ \hline
			
			-- & -- & \checkmark 
			& -- 
			& -- 
			& \textbf{32.68}/\textbf{0.959} \\ \hline
			
			\checkmark & \checkmark & -- 
			& 31.72/0.916 \; / \; 28.57/0.822
			& 36.29/0.971
			& -- \\ \hline
			
			\checkmark & -- & \checkmark 
			& 31.70/0.915 \; / \; 28.52/0.821
			& -- 
			& 32.44/0.955 \\ \hline
			
			-- & \checkmark & \checkmark 
			& -- 
			& 36.24/0.970
			& 32.39/0.954 \\ \hline
			
			\checkmark & \checkmark & \checkmark 
			& 31.56/0.913 \; / \; 28.33/0.845
			& 35.21/0.958
			& 31.34/0.914 \\ \hline
			
	\end{tabular}}
\end{table}
Figure~\ref{fig:pareto} visualises the parameter--accuracy trade-off. MDIRNET occupies the top-left region of the plot, achieving the highest average PSNR among the four methods while requiring 13-38$\times$ fewer parameters than AirNet, MPRNet, and Restormer. This provides empirical evidence that the model-driven DU-OVPCA core, which performs restoration via algorithmic SVD operations rather than parameterized convolutions, achieves strong accuracy at a fraction of the model size.\\
Table~\ref{tab:ablation} shows the effect of different degradation combinations when using MDIRNET. Across the evaluated combinations, the proposed model maintains strong performance, with PSNR values ranging from 28 to 36.5~dB depending on the degradation combination. When all degradations are combined, MDIRNET achieves 31.56~dB / 0.913 on BSD68 ($\sigma=25$) and 35.21~dB / 0.958 on Rain100L, supporting its robustness to complex degradation mixtures.\\
The results in Table~\ref{table4} highlight two main observations. In the one-by-one setting, MDIRNET achieves strong PSNR and SSIM across BSD68, Rain100L, and GoPro with a runtime of 0.19--0.23~s and 0.68~M parameters / 14.0~GFLOPs, substantially fewer than AirNet (8.93~M, 75.6~GFLOPS) and Restormer (26.13~M, 141.0~GFLOPS).\\
The One-By-One and All-In-One results follow the provenance and inference settings in Table~\ref{tab:baseline_protocol}. Literature results are not treated as retrained unless stated; differences may reflect training data, degradation settings, evaluation protocols, or sources.
Since methods such as ISTA-Net+ and ADMM-Net were originally designed for compressive sensing rather than multi-degradation restoration, they are not included as primary baselines in the main experimental comparison. ISTA-Net+ (31.04 dB, 0.34M) and ADMM-Net (30.91 dB, 0.42M) outperform the 0.02M-parameter DU-OVPCA (30.67 dB), reflecting their role as lightweight, model-driven backbones rather than standalone restoration networks. The full MDIRNET reaches 33.68 dB. Nevertheless, DU-OVPCA shares the same unfolding principle and is further evaluated through the layer-depth ablation (Table~\ref{tab:layer_ablation}, Fig.~\ref{fig:combined_analysis}).
\paragraph{Noise Degradation.}
Figure~\ref{fig:single_noise} compares denoising results. Baseline methods such as DnCNN and BRDNet oversmooth high-frequency textures, whereas MDIRNET produces sharper outputs with better PSNR/SSIM, balancing noise suppression and detail preservation.
\paragraph{Rain Degradation.}
Figure~\ref{fig:single_rain} shows that UMRL and SIRR leave residual streaks, while MDIRNET removes most rain while preserving sharp boundaries and textures.
\paragraph{Blur Degradation.}
Figure~\ref{fig:single_blur} shows that DeblurGAN and MPRNet introduce ringing or oversharpening, whereas MDIRNET restores edges and textures with minimal distortion.\\
Across all single degradations, MDIRNET effectively removes corruptions while preserving structural details. Unlike competing methods that either oversmooth or introduce artifacts, MDIRNET achieves competitive PSNR/SSIM scores, demonstrating strong adaptability and perceptual fidelity.
	\subsection{Qualitative Results on Multiple Degradations}
	Figure~\ref{fig:qual_all} presents multi-degradation comparisons.
	\textcolor{red}{
	\begin{figure*}[t]
		\centering
		\includegraphics[width=\textwidth]{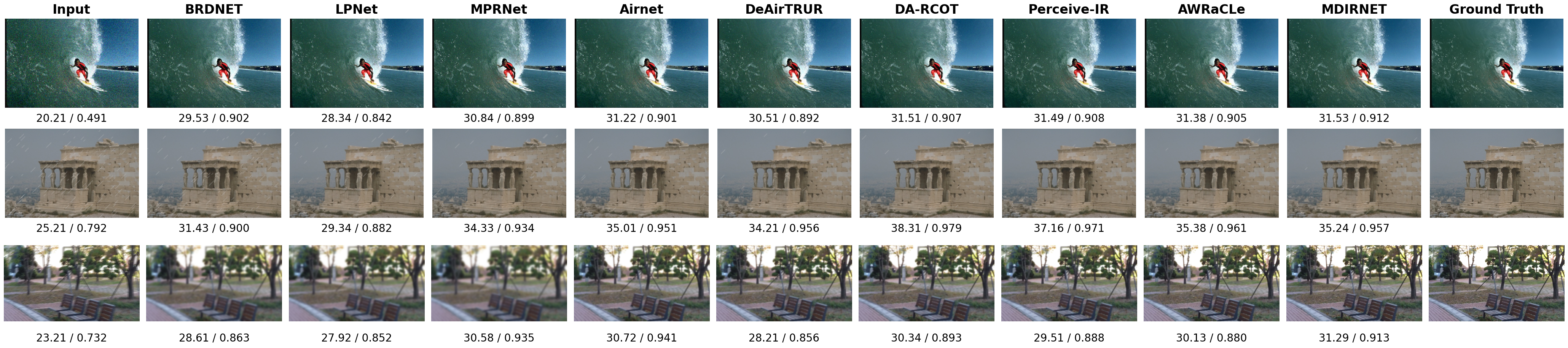}
		\caption{Qualitative comparison across noise, rain, and blur. MDIRNET achieves cleaner restorations with better texture and structural preservation.}
		\label{fig:qual_all}
	\end{figure*}}
This comparison illustrates MDIRNET's performance across the evaluated degradation settings. Compared with the baseline methods, MDIRNET produces cleaner restorations while preserving structural details under the evaluated benchmark conditions.

\subsection{Comparison with Alternative Deep-Unfolded Backbones}
To evaluate DU-OVPCA, we compare it with unfolded DU-RPCA~\cite{imran2022deep}, DU-WNNM~\cite{yair2018multi}, and a parameter-matched CNN under identical training settings. The CNN core replaces DU-OVPCA with a lightweight two-layer residual block (3$\times$3, 32 channels, ReLU) operating on the patch-group tensor $\mathbf{D}_{\mathcal{N}_i}$ via shared weights across the $T$ patches, preserving cross-patch interaction without low-rank or SVD constraints. DRAM is enabled only for MDIRNET. We use a custom minimal CNN to ensure a controlled comparison. Unlike pre-trained models, it avoids training-data bias and capacity mismatch. Both the CNN and DU-OVPCA are trained on the same data, with the same loss and hyperparameters, thereby isolating the effect of architecture.\\
\begin{table*}[t]
		\centering
		\caption{Comparison of deep-unfolded low-rank backbones and a parameter-matched CNN under identical training settings.}
		\label{tab:backbone}
		\renewcommand{\arraystretch}{1.15}
		
		\begin{tabular}{|l|c|c|c|c|c|c|c|}
			\hline
			\textbf{Method} &
			\textbf{Low-rank Prior} &
			\makecell{\textbf{Deep}\\\textbf{Unfolding}} &
			\makecell{\textbf{DRAM}} &
			\textbf{BSD68 ($\sigma = 25$)} &
			\textbf{Rain100L} &
			\textbf{GoPro} &
			\textbf{Params (M)} \\
			\hline
			
			CNN core (PPM+SAM) &
			$\times$ &
			$\times$ &
			$\times$ &
			29.85 &
			32.41 &
			29.77 &
			0.67 \\
			\hline
			
			DU-WNNM backbone (PPM+SAM) &
			$\checkmark$ &
			$\checkmark$ &
			$\times$ &
			30.71 &
			33.48 &
			30.62 &
			0.72 \\
			\hline
			
			DU-RPCA backbone (PPM+SAM) &
			$\checkmark$ &
			$\checkmark$ &
			$\times$ &
			30.96 &
			33.55 &
			30.87 &
			0.85 \\
			\hline
			
			DU-OVPCA (PPM+SAM, NO DRAM) &
			$\checkmark$ &
			$\checkmark$ &
			$\times$ &
			31.29 &
			34.75 &
			31.21 &
			0.67 \\
			\hline
			
			\textbf{MDIRNET (DU-OVPCA+DRAM)} &
			$\checkmark$ &
			$\checkmark$ &
			$\checkmark$ &
			\textbf{31.84} &
			\textbf{36.53} &
			\textbf{32.68} &
			0.68 \\
			\hline
			
		\end{tabular}
\end{table*}
This experiment isolates the restoration backbone under a controlled protocol, with PPM and SAM fixed and DRAM disabled. It evaluates the restoration core without claiming universal backbone superiority. As shown in Table~\ref{tab:backbone}, replacing DU-OVPCA with a parameter-matched CNN core reduces PSNR by 1.44 dB (BSD68), 2.34 dB (Rain100L), and 1.44 dB (GoPro), despite identical parameter count, showing that the DU-OVPCA restoration core provides a substantial performance contribution beyond the shared PPM and SAM components. DU-OVPCA also consistently outperforms DU-WNNM and DU-RPCA, while DRAM provides additional gains of +0.55, +1.78, and +1.47 dB on BSD68, Rain100L, and GoPro, respectively.

\section{Ablation Study} 
\begin{figure*}[!htbp]
	\centering
	\subfloat[]{
		\includegraphics[
		height=4cm,
		trim=25 2 25 2,
		clip
		]{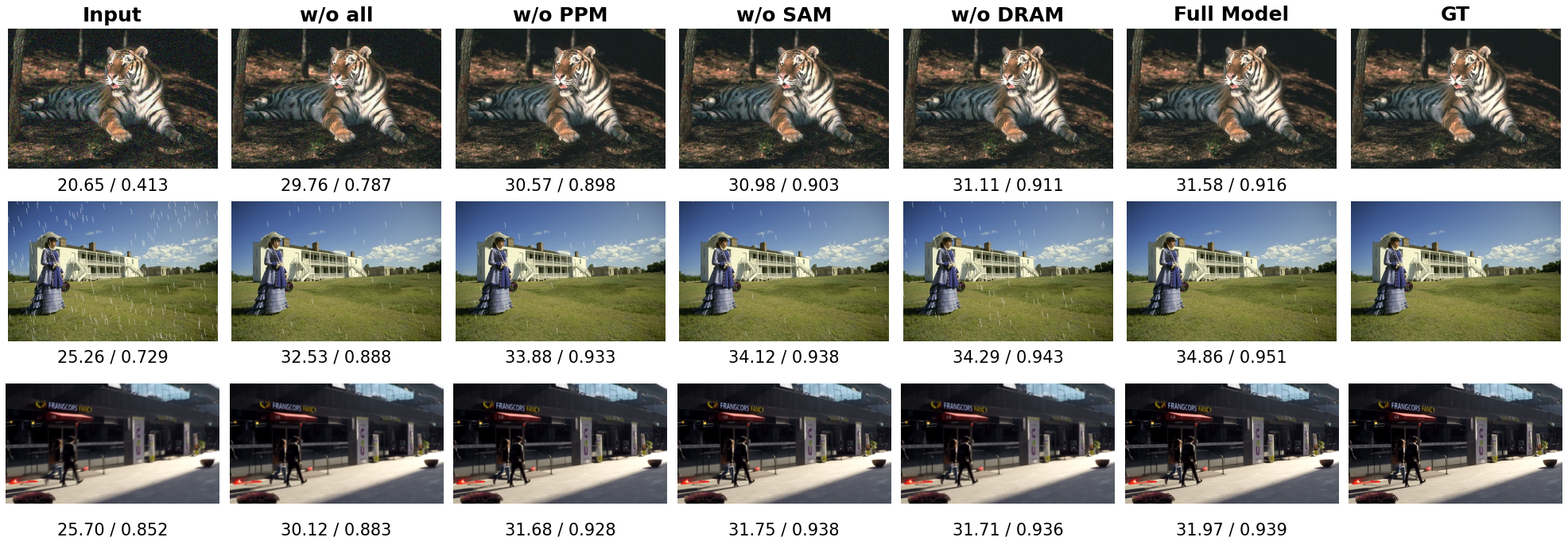}
		\label{fig:ablation-mdirnet}
	}
	\hspace{0.5mm}
	\subfloat[]{
		\includegraphics[
		height=4.7cm,
		trim=25 2 25 2,
		clip
		]{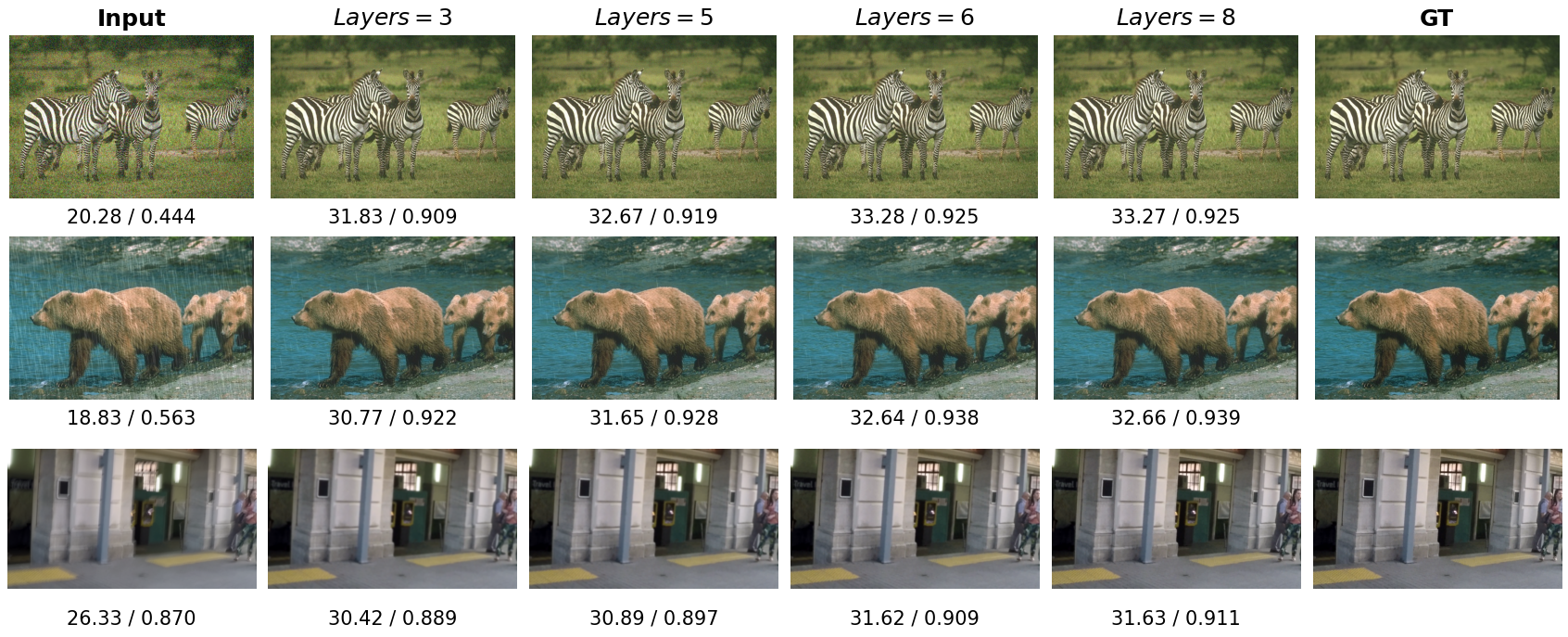}
		\label{fig:dvupca-layer}
	}
	\hspace{0.5mm}
	\subfloat[]{
		\includegraphics[
		height=4.7cm,
		trim=25 2 25 2,
		clip
		]{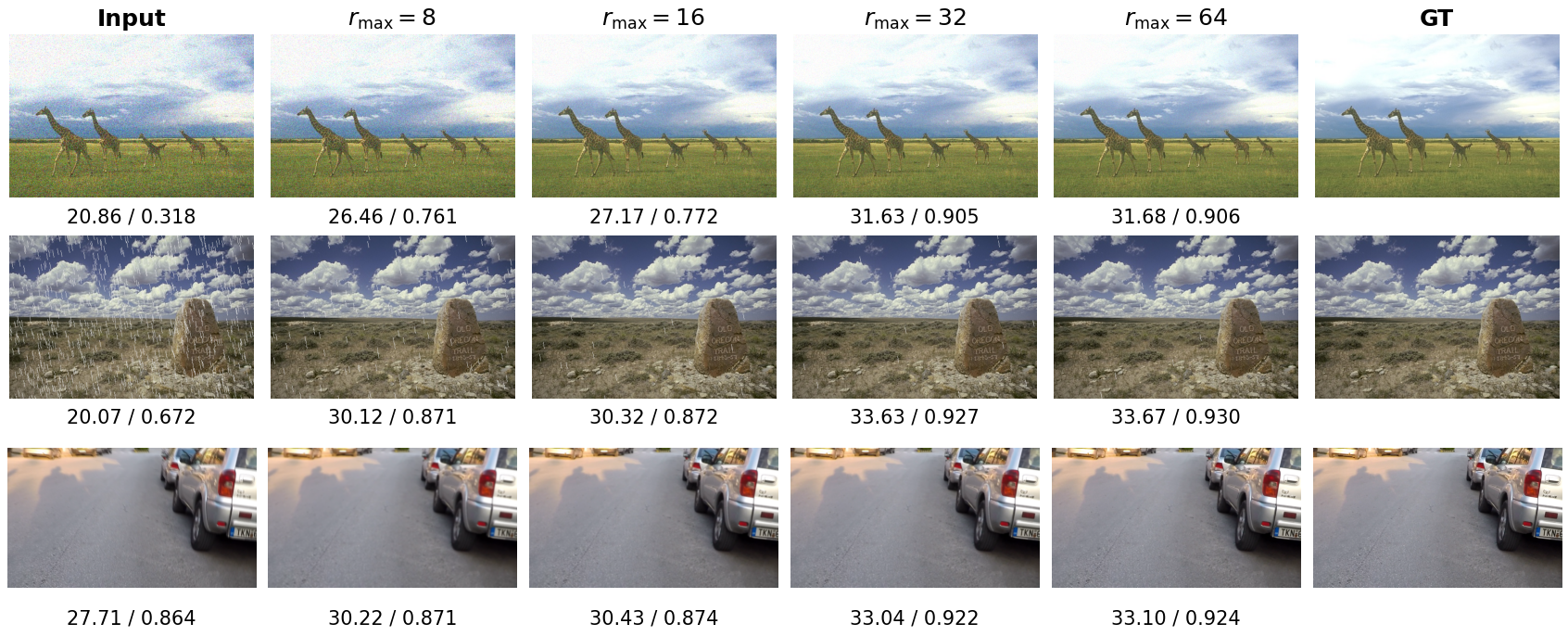}
		\label{fig:dvupca-rank}
	}
	\caption{Quantitative ablation analysis. (a)~Effect of removing SAM, PPM, and DRAM. (b)~Performance versus DU-OVPCA depth $K$. (c)~Performance versus maximum rank $r_{\mathrm{max}}$.}
	\label{fig:ablation-all}
\end{figure*}

\begin{figure*}[!htbp]
	\centering
	\subfloat[PPM patch-centre distribution]{
		\includegraphics[height=4.0cm]{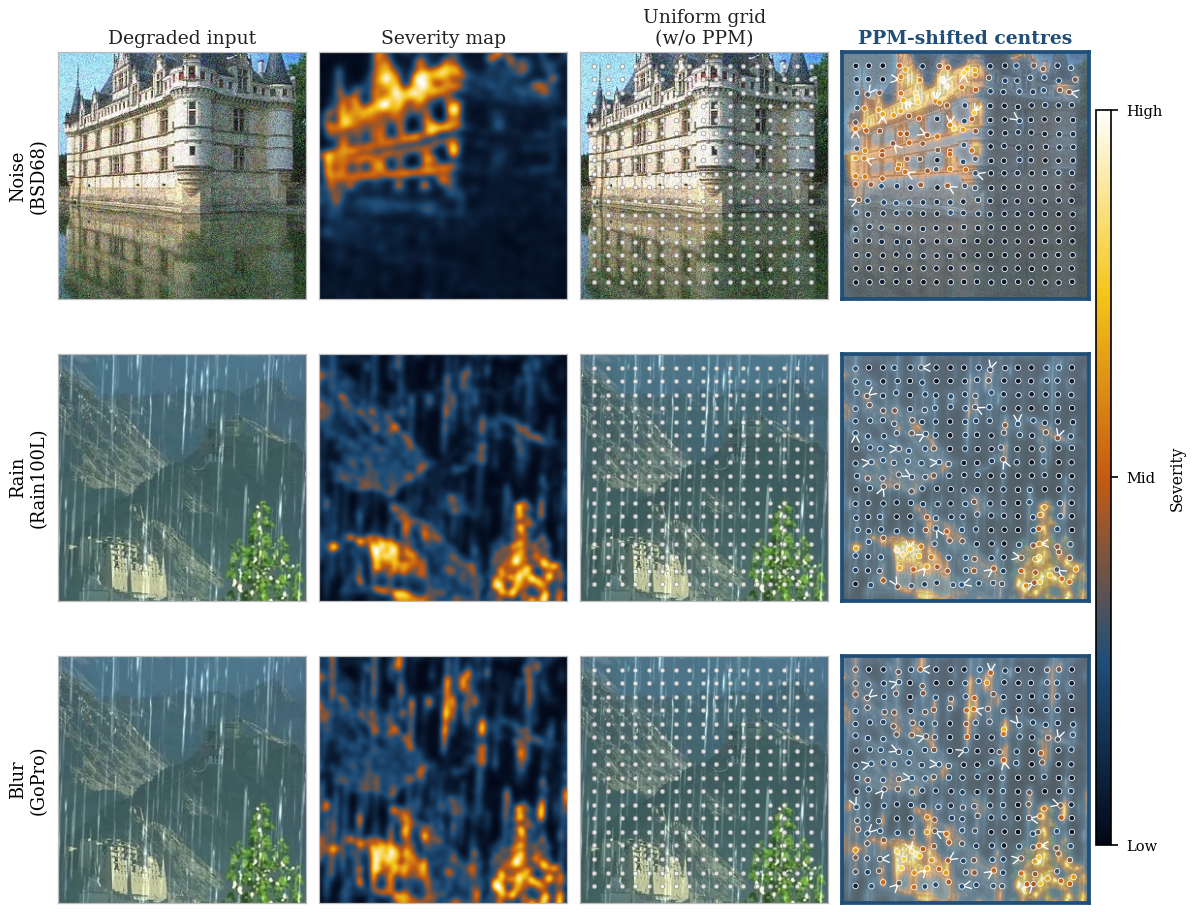}
		\label{fig:ppm-vis}
	}
	\hfill
	\subfloat[DRAM rank allocation maps]{
		\includegraphics[height=4.0cm]{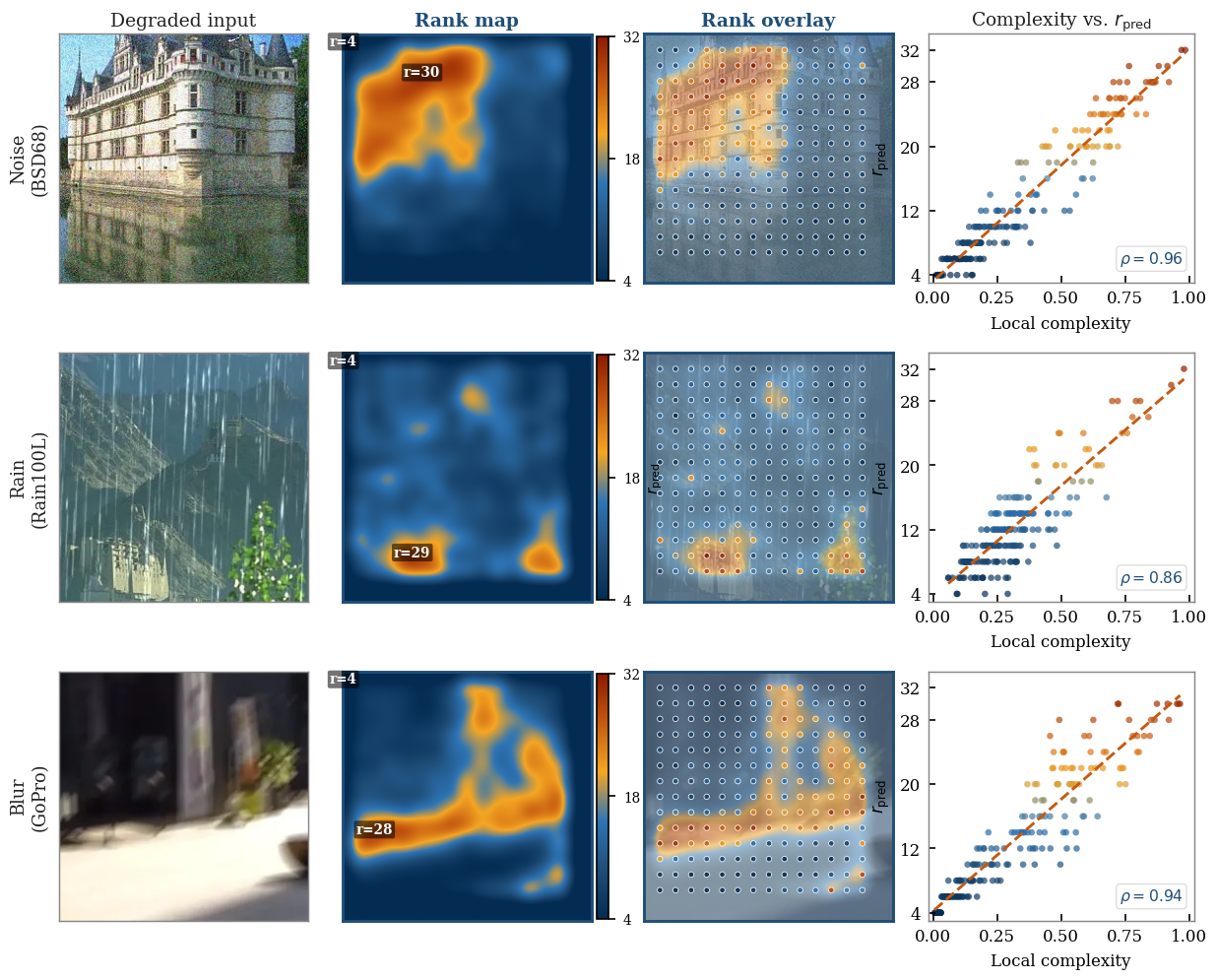}
		\label{fig:dram-rank}
	}
	\hfill
	\subfloat[SAM attention (gate) maps]{
		\includegraphics[height=4.0cm]{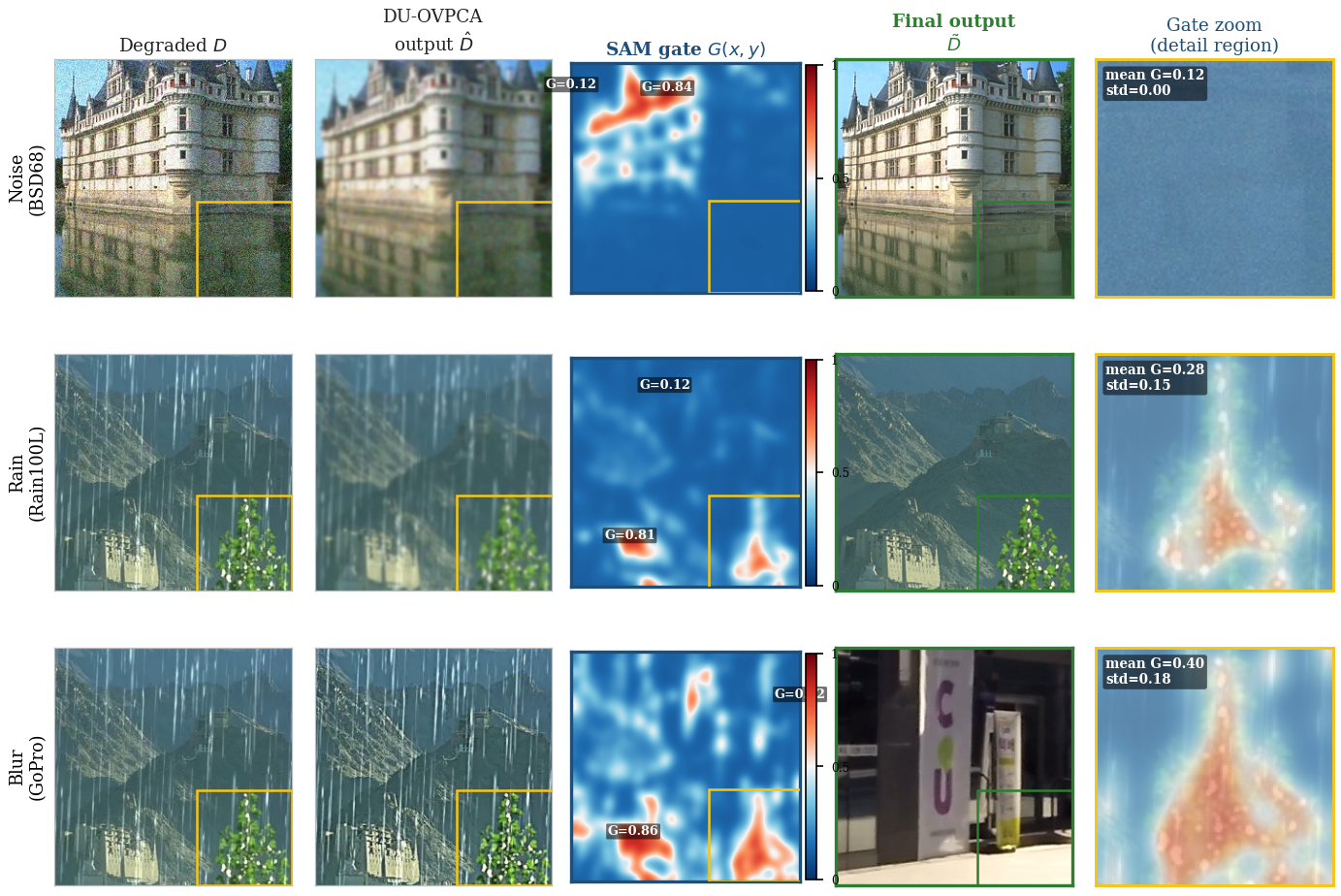}
		\label{fig:sam-gate}
	}
\caption{Mechanism validation. (a)~PPM: learned patch centers show spatially adaptive sampling. (b)~DRAM: higher ranks in challenging regions ($\rho=0.87$--$0.96$). (c)~SAM: lower gates in well-restored regions and higher gates in oversmoothed regions.}
	\label{fig:mechanism-vis}
\end{figure*}

 Tables~\ref{tab:ablation-quantitative}--\ref{tab:lambda_unc}, Fig.~\ref{fig:ablation-all}, and Fig.~\ref{fig:mechanism-vis} summarise the ablation results on BSD68 ($\sigma{=}25$), Rain100L, and GoPro.

\begin{table}[ht]
	\centering
	\caption{MDIRNET ablation results, including module ablation, PPM sampling, DRAM feature aggregation, and reconstruction losses. BSD68 results are reported at $\sigma=25$.}
	\label{tab:ablation-quantitative}
	\setlength{\tabcolsep}{3pt}
	\resizebox{\linewidth}{!}{%
		\begin{tabular}{|l|l|c|c|c|c|c|}
			\hline
			\textbf{Group} & \textbf{Variant} & \textbf{BSD68} &
			\textbf{Rain100L} & \textbf{GoPro} & \textbf{Time (s)} &
			\textbf{Avg. PSNR/SSIM} \\
			\hline
			
			\multirow{5}{*}{\textit{Modules}}
			& MDIRNET (Full)
			& \textbf{31.84/0.920}
			& \textbf{36.53/0.974}
			& \textbf{32.68/0.959}
			& 0.195
			& \textbf{33.68/0.951} \\
			
			& w/o SAM
			& 31.39/0.917
			& 34.90/0.931
			& 31.14/0.934
			& 0.218
			& 32.48/0.927 \\
			
			& w/o DRAM (fixed $r$=32)
			& 31.29/0.915
			& 34.75/0.929
			& 31.21/0.934
			& 0.351
			& 32.42/0.926 \\
			
			& w/o PPM (fixed grid)
			& 31.02/0.909
			& 34.34/0.927
			& 31.19/0.932
			& 0.225
			& 32.18/0.923 \\
			
			& Baseline DU-OVPCA
			& 29.46/0.802
			& 32.96/0.873
			& 29.60/0.879
			& 0.348
			& 30.67/0.851 \\
			\hline
			
			\multirow{5}{*}{\textit{PPM sampling}}
			
			& Fixed grid
			& 31.02/0.909
			& 34.34/0.927
			& 31.19/0.932
			& 0.225
			& 32.18/0.923 \\
			
			& Gradient-based
			& 31.15/0.911
			& 35.10/0.943
			& 31.80/0.941
			& 0.205
			& 32.68/0.932 \\
			
			& Smoothed gradient
			& 31.12/0.911
			& 35.05/0.941
			& 31.78/0.940
			& 0.208
			& 32.65/0.931 \\
			
			& Random adaptive
			& 31.08/0.910
			& 34.92/0.937
			& 31.45/0.936
			& 0.201
			& 32.48/0.928 \\
			
			& \textbf{Learned PPM}
			& \textbf{31.84/0.920}
			& \textbf{36.53/0.974}
			& \textbf{32.68/0.959}
			& \textbf{0.195}
			& \textbf{33.68/0.951} \\
			\hline
			
			\multirow{4}{*}{\textit{DRAM aggregation.}}
			
			& Fixed $r$=32 (no DRAM)
			& 31.29/0.915
			& 34.75/0.929
			& 31.21/0.934
			& 0.351
			& 32.42/0.926 \\
			
			& Max pooling
			& 31.57/0.917
			& 35.90/0.961
			& 32.20/0.945
			& 0.200
			& 33.22/0.941 \\
			
			& Attention-weighted
			& 31.81/0.919
			& 36.45/0.972
			& 32.61/0.957
			& 0.228
			& 33.62/0.949 \\
			
			& \textbf{Mean (MDIRNET)}
			& \textbf{31.84/0.920}
			& \textbf{36.53/0.974}
			& \textbf{32.68/0.959}
			& \textbf{0.195}
			& \textbf{33.68/0.951} \\
			\hline
			
			\multirow{3}{*}{\textit{Loss fn.}}
			
			& L1 only
			& 31.57/0.916
			& 36.10/0.969
			& 32.30/0.952
			& 0.223
			& 33.32/0.946 \\
			
			& \textbf{L1+L2 (MDIRNET)}
			& \textbf{31.84/0.920}
			& \textbf{36.53/0.974}
			& \textbf{32.68/0.959}
			& 0.195
			& \textbf{33.68/0.951} \\
			
			& L1+SSIM
			& 31.63/0.923
			& 36.22/0.976
			& 32.45/0.963
			& 0.225
			& 33.43/0.954 \\
			\hline
			
	\end{tabular}}
\end{table}
Besides validating the proposed modules, Table~\ref{tab:ablation-quantitative} compares DRAM aggregation strategies. Mean pooling achieves the highest accuracy with the lowest measured inference time, outperforming max and attention-weighted aggregation. It is therefore adopted throughout MDIRNET.
\begin{table}[h]
	\centering
	\caption{Ablation of maximum rank ($r_{\max}$) in MDIRNET. Entries are PSNR/SSIM; last two columns show inference time (s) and average PSNR/SSIM.}
	\label{tab:rank_ablation}
	\setlength{\tabcolsep}{3pt}
	
	\resizebox{\columnwidth}{!}{%
		\begin{tabular}{|c|c|c|c|c|c|}
			\hline
			\textbf{$r_{max}$} &
			\textbf{BSD68} &
			\textbf{GoPro} &
			\textbf{Rain100L} &
			\textbf{Time (s)} &
			\textbf{Avg.} \\
			\hline
			8  & 26.59/0.765 & 28.39/0.843 & 32.55/0.880 & 0.082 & 29.18/0.829 \\
			\hline
			16 & 27.29/0.785 & 28.79/0.858 & 32.85/0.886 & 0.151 & 29.64/0.843 \\
			\hline
			\textbf{32} &
			\textbf{31.84}/\textbf{0.920} &
			\textbf{32.68}/\textbf{0.959} &
			\textbf{36.53}/\textbf{0.974} &
			0.195 &
			\textbf{33.68}/\textbf{0.951} \\
			\hline
			64 & 31.80/0.919 & 32.60/0.956 & 36.40/0.972 & 0.336 & 33.60/0.949 \\
			\hline
		\end{tabular}%
	}
\end{table}

\begin{table}[t]

	\centering
	\caption{Region-wise DRAM rank allocation across image characteristics.}
	\label{tab:dram_rank_distribution}
	\setlength{\tabcolsep}{4pt}
	
	\begin{tabular}{|l|c|c|}
		\hline
		\textbf{Region Type} &
		\textbf{Mean Rank} &
		\textbf{PSNR (dB)} \\
		\hline
		
		Smooth ($g<p_{25}$) &
		$8.2 \pm 1.1$ &
		40.1 \\
		\hline
		
		Textured ($p_{25}\leq g\leq p_{75}$) &
		$14.6 \pm 2.3$ &
		34.8 \\
		\hline
		
		High-gradient edges ($g>p_{75}$) &
		$22.4 \pm 3.1$ &
		29.5 \\
		\hline
		
	\end{tabular}
	
	\vspace{1mm}
	
	\begin{minipage}{\columnwidth}
		\footnotesize
		\textit{Note:} $g$ denotes local gradient magnitude; $p_{25}$ and $p_{75}$ denote its 25th and 75th percentiles. Results are averaged across BSD68, Rain100L, and GoPro.
	\end{minipage}
	
\end{table}

\begin{table}[h]
	\centering
	\caption{Ablation of DU-OVPCA unfolding depth. BSD68 results are for $\sigma=25$. Entries are PSNR/SSIM; the last two columns report inference time (s) and average PSNR/SSIM.}
	\label{tab:layer_ablation}
	\resizebox{\columnwidth}{!}{%
		\begin{tabular}{|c|c|c|c|c|c|}
			\hline
			\textbf{Layers} & \textbf{BSD68} & \textbf{GoPro} & \textbf{Rain100L} & \textbf{Avg. Time (s)} & \textbf{Avg. PSNR/SSIM} \\
			\hline
			3 & 30.42/0.908 & 30.39/0.916 & 33.60/0.918 & 0.190 & 31.47/0.914 \\
			\hline
			5 & 31.27/0.917 & 30.99/0.927 & 34.43/0.930 & 0.193 & 32.23/0.918 \\
			\hline
			\textbf{6}
			& \textbf{31.84}/\textbf{0.920}
			& \textbf{32.68}/\textbf{0.959}
			& \textbf{36.53}/\textbf{0.974}
			& 0.195
			& \textbf{33.68}/\textbf{0.951} \\
			\hline
			8 & 31.77/0.917 & 32.60/0.956 & 36.40/0.972 & 0.205 & 33.59/0.948 \\
			\hline
		\end{tabular}%
	}
\end{table}

\begin{figure*}[t!]
	\centering
	\subfloat[]{\includegraphics[width=0.25\textwidth]{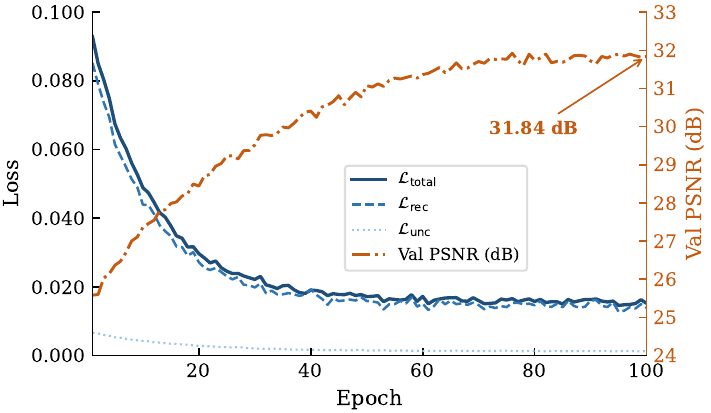}\label{fig:convergence-train}}
	\hfill
	\subfloat[]{\includegraphics[width=0.25\textwidth]{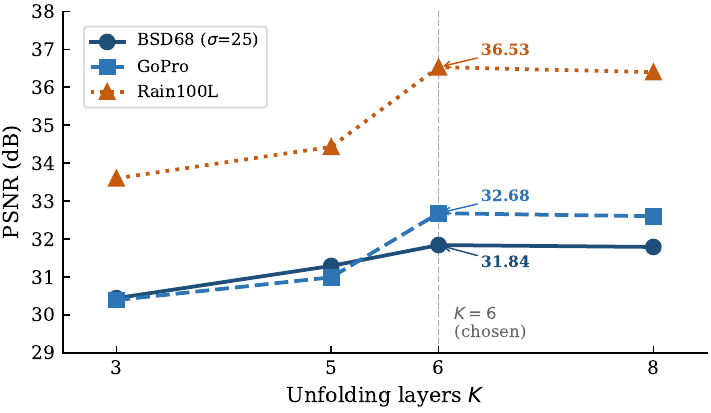}\label{fig:convergence-layer}}
	\hfill
	\subfloat[]{\includegraphics[width=0.25\textwidth]{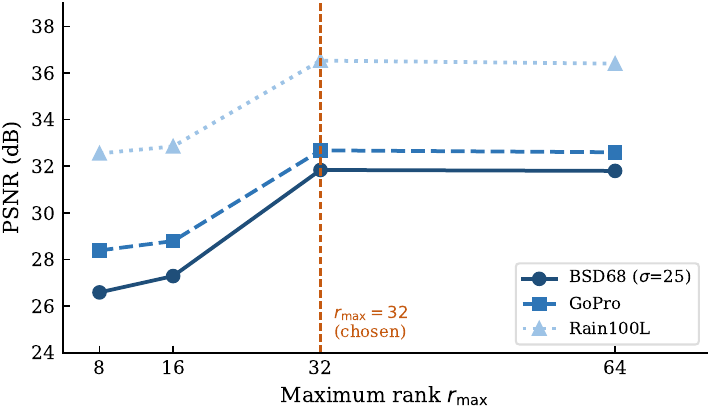}\label{fig:rmax-ablation}}
	\caption{Training and ablation analysis. (a)~Training-loss evolution over 100 epochs. (b)~Per-layer PSNR; saturation at $K{=}6$. (c)~$r_\mathrm{max}$ ablation; saturation at 32.}
	\label{fig:combined_analysis}
\end{figure*}

\textbf{Mechanism Validation:} Learned PPM achieves a higher average PSNR than gradient-based sampling (33.68 vs. 32.68~dB) and remains stable across five seeds ($<$0.13~dB standard deviation), providing empirical evidence for the effectiveness of learned spatially adaptive sampling under the evaluated settings. DRAM assigns higher ranks to textured/challenging regions, indicating content-dependent allocation. SAM improves all evaluated regions, with varying gate values. These results support the observed roles of PPM, DRAM, and SAM, but do not establish estimates of degradation severity or causal mechanisms.\\
\textbf{Effect of PPM, DRAM, and SAM:} For PPM, Disabling learnable patch partitioning reduces PSNR by 0.82~dB on BSD68, 2.19~dB on Rain100L, and 1.49~dB on GoPro, indicating the empirical benefit of content-adaptive patch allocation under the evaluated settings. Adaptive rank selection further improves restoration quality and reduces measured runtime, without explicitly optimizing runtime during training.\\
Removing SAM drops PSNR by 0.45~dB on BSD68, 1.63~dB on Rain100L, and 1.54~dB on GoPro, with visible residual artifacts. SAM adds a modest computational cost in terms of GFLOPs while consistently improving both PSNR and SSIM.\\
\textbf{Maximum Rank $r_{max}$ in DRAM:} We also tested different $r_{\max}$ values (Table \ref{tab:rank_ablation}). Low rank limits ($r_{\max}=16$) reduced PSNR by about 4.0 dB, while higher values gave minimal gains but increased runtime. Overall, $r_{\max}=32$ achieved the best balance between accuracy and efficiency.\\
\textbf{Unfolding Depth:} We also evaluated different numbers of DU-OVPCA layers $K$ (Table \ref{tab:layer_ablation}). Increasing $K$ from 3 to 6 improves restoration quality, while $K=8$ yields negligible gains at higher cost. Thus, $K=6$ offers a practical accuracy--complexity trade-off.\\
Figure~\ref{fig:combined_analysis} presents the training-loss evolution and ablation analyses, together with the observed saturation at $K{=}6$ layers and $r_\mathrm{max}{=}32$.\\
	\textbf{Consistency Regularizer Weight:}
	Table~\ref{tab:lambda_unc} shows the effect of the group-wise consistency regularizer weight $\lambda_{\text{unc}}$. Increasing $\lambda_{\text{unc}}$ from 0 to 0.1 improves restoration performance, while larger values degrade it. Therefore, $\lambda_{\text{unc}}=0.1$ is adopted throughout the paper. Moderate regularization ($\lambda_{\mathrm{unc}}=0.1$) improves both PSNR (+0.53dB on BSD68) and LPIPS (0.052$\rightarrow$0.049), indicating better perceptual quality without oversmoothing. Stronger regularization degrades both metrics, indicating that moderate regularization is optimal.
\begin{table}[h]
	\centering
	\caption{Ablation study on consistency regularizer weight $\lambda_{\text{unc}}$. 
		PSNR and LPIPS both improve at $\lambda_{\text{unc}}=0.1$, confirming enhanced 
		perceptual quality without oversmoothing.}
	\label{tab:lambda_unc}
	\setlength{\tabcolsep}{3pt}
	
	\begin{tabular}{|c|c|c|c|c|c|c|}
		\hline
		$\lambda_{\text{unc}}$ & BSD68 & LPIPS & GoPro & LPIPS & 
		Rain100L & LPIPS \\
		\hline
		0      & 31.31 & 0.052 & 31.79 & 0.048 & 35.80 & 0.031 \\
		\hline
		0.01   & 31.42 & 0.052 & 32.00 & 0.047 & 35.91 & 0.030 \\
		\hline
		0.05   & 31.65 & 0.051 & 32.32 & 0.045 & 36.22 & 0.029 \\
		\hline
		\textbf{0.1} & \textbf{31.84} & \textbf{0.049} & \textbf{32.68} & 
		\textbf{0.042} & \textbf{36.53} & \textbf{0.028} \\
		\hline
		0.2    & 31.60 & 0.051 & 32.33 & 0.046 & 36.18 & 0.029 \\
		\hline
		0.5    & 31.28 & 0.054 & 31.92 & 0.049 & 35.90 & 0.032 \\
		\hline
	\end{tabular}
	
\end{table}

\section{Conclusion}
In this work, we presented MDIRNET, a unified deep unfolding network for multi-degradation image restoration. Built on an OVPCA-based core, it combines model-driven optimization with deep learning. We incorporated three architectural modules: a learnable Patch Partitioning Module (PPM) for content-aware sampling of image patches, a DRAM for content-adaptive low-rank modeling, and a Supervised Attention Module (SAM) for residual refinement. Extensive experiments demonstrate that MDIRNET achieves competitive performance across denoising, deraining, and deblurring while providing a strong efficiency--accuracy trade-off within a unified framework. The controlled backbone comparison further shows that DU-OVPCA provides a substantial performance benefit under the evaluated settings. Nevertheless, several limitations remain. All runtime measurements were conducted on an NVIDIA RTX~3090; CPU, edge, and embedded deployment efficiency, peak memory, and energy consumption remain unvalidated. The low-rank prior is less effective for severe large-kernel blur, and validation on real-world mixed-degradation datasets (e.g., DND) remains future work. Fixed patch size and spatial grouping may also limit adaptability across image structures and degradation scales. Future work will investigate adaptive multi-scale patch partitioning, explicit non-local patch grouping, lightweight deployment on edge devices, and formal convergence analysis of the end-to-end unfolded network.
\balance
\bibliographystyle{IEEEtran}
\bibliography{main}

\end{document}